\documentclass[aps,prl,floats,twocolumn,showpacs,superscriptaddress]{revtex4-1}

\usepackage{graphicx}
\usepackage{amssymb}
\usepackage{color}

\begin{document}
\title{Persistence and Stochastic Periodicity in the Intensity Dynamics of a Fiber Laser During the Transition to Optical Turbulence}

\author{Laura Carpi}

\affiliation{Programa de P\'os-Gradua\c c\~ao em Modelagem Matem\'atica e Computacional, PPGMMC, Centro Federal de Educa\c c\~ao Tecnol\'ogica de Minas Gerais, CEFET-MG.  Av. Amazonas, 7675. 30510-000. Belo Horizonte, MG, Brazil}
\author{Cristina Masoller}
\email{cristina.masoller@upc.edu} 
\affiliation{Departament de Fisica, Universitat Politecnica de Catalunya, Rambla Sant Nebridi 22, ES-08222 Terrassa, Barcelona, Spain}

\begin{abstract}
Many natural systems display transitions among different dynamical regimes, which are difficult to identify when the data is noisy and high dimensional. A technologically relevant example is a fiber laser, which can display complex dynamical behaviors that involve nonlinear interactions of millions of cavity modes. Here we study the laminar--turbulence transition that occurs when the laser pump power is increased. By applying various data analysis tools to empirical intensity time series we characterize their persistence and demonstrate that at the transition temporal correlations can be precisely represented by a surprisingly simple model. 
\end{abstract}
\maketitle
\section{Introduction}
Fibre lasers are technologically relevant laser systems that can display complex spatio-temporal dynamics which involve nonlinear interactions of a huge number of cavity modes \cite{Turitsyn2010,sergei_2017,real_time_2017}. The transition to ``optical turbulence'' as the laser pump power is increased has received attention and experimental observations of spatio-temporal dynamical regimes have yield new light into the rich underlying nonlinear physics \cite{natphot,natcom_2015,dima_2016,prl_2016}. 

In \cite{natphot,natcom_2015} the laser output intensity was investigated by exploiting the analogy with an spatially extended system: the evolution of the intensity during one cavity round-trip time occurs in a ``space-like dimension'', while the evolution during many round-trips, occurs in a ``temporal dimension''. This two-dimensional representation of the output of dynamical systems that have a well-defined characteristic time-scale (such as the cavity round-trip time) has proven to be very useful to uncover hidden space-like features, such as defects and dislocations \cite{arecchi_1992,gianni_1996}.

In the fiber laser, long-range correlations from one round-trip to the next (i.e., in the temporal dimension) have been identified \cite{natphot,natcom_2015}, and during the laminar-turbulence transition, shorter correlations (in the space-like dimension) have also been detected \cite{prl_2016}. They were identified with two time series analysis methods: the horizontal visibility graph (HVG) \cite{Lacasa2009, HVG} and ordinal analysis \cite{bandt_PRL_2002,review_rosso,review_amigo,small_2014} (described in \textit{Appendix A}).

The HVG method maps a time series into a graph that keeps information about the temporal ordering of the data points in the time series. In~\cite{prl_2016} the graph was characterized by Shannon entropy, $S[P]$, computed from the degree distribution, $P(k)$, that gives the probability that a node (i.e., a data point $I_i$) has $k$ links. $S[P]$ (referred to as HVG entropy) was computed for various laser pump powers. When the intensity time series were pre-processed such that only the intensity peaks higher than a threshold were analyzed, a sharp decrease of the HVG entropy was detected at the transition. In contrast, the HVG entropy decreased smoothly when all the data points were analyzed. Ordinal analysis transforms a time series into a sequence of symbols (ordinal patterns, OPs), also keeping the information about the temporal ordering of the data points in the time series. The OPs sequence is then characterized by the ordinal probabilities and the associated entropy (permutation entropy, PE). In~\cite{prl_2016} it was shown that the PE displays, as the pump power increases, the same variation as the HVG entropy. In addition, by using a lag-time to define the OPs, short scale correlations (in the space-like dimension) were detected.

Temporally correlated signals have been observed in many fields (examples include EEG signals, stock market prices, water flows through rivers, earthquake inter-event intervals, rainfall and climatic time series, among many others). A lot of work has been devoted to characterize them (typically the \textit{persistence}, measured by the Hurst exponent, $\mathcal{H}$ \cite{Mandelbrot_1968}), and to understand the physical mechanisms underlying such correlations.

Here we re-analyze the fiber laser empirical data studied in \cite{prl_2016} to address both issues.  In the first part, by using the HVG method, we identify synthetic data with similar short correlations as the laser data. Specifically, we consider a stochastic processes with known $\mathcal{H}$: \textit{fractional Brownian motion} (fBm) \cite{Mandelbrot_1968,Muzy1991,Stolovitzky1994,perez_josa_2004,bio_2013,olivares_2016,npg_2017}. When $\mathcal{H} > 0.5$ consecutive increments tend to have the same sign and the fBm process is \textit{persistent}; in contrast, when $\mathcal{H} < 0.5$ consecutive increments tend to have opposite signs, and the process is \textit{anti-persistent}. In the second part of our work we demonstrate, by using ordinal analysis, that a very simple model accounts for the correlations among lagged intensity values that were identified in~\cite{prl_2016}.

\section{Experimental setup and datasets}
 
The experimental setup and datasets are described in~\cite{prl_2016}: the laser is a Raman fiber (normal dispersion) of 1 km placed between two fiber Bragg gratings acting as cavity mirrors; the pump power is varied from below to above the transition (which occurs for 0.9 W), and for each pump power a time series with $5\times 10^7$ data points was recorded, with resolution $dt=0.0125$~ns.  

\begin{figure}
\centering
\includegraphics[width=0.79\columnwidth]{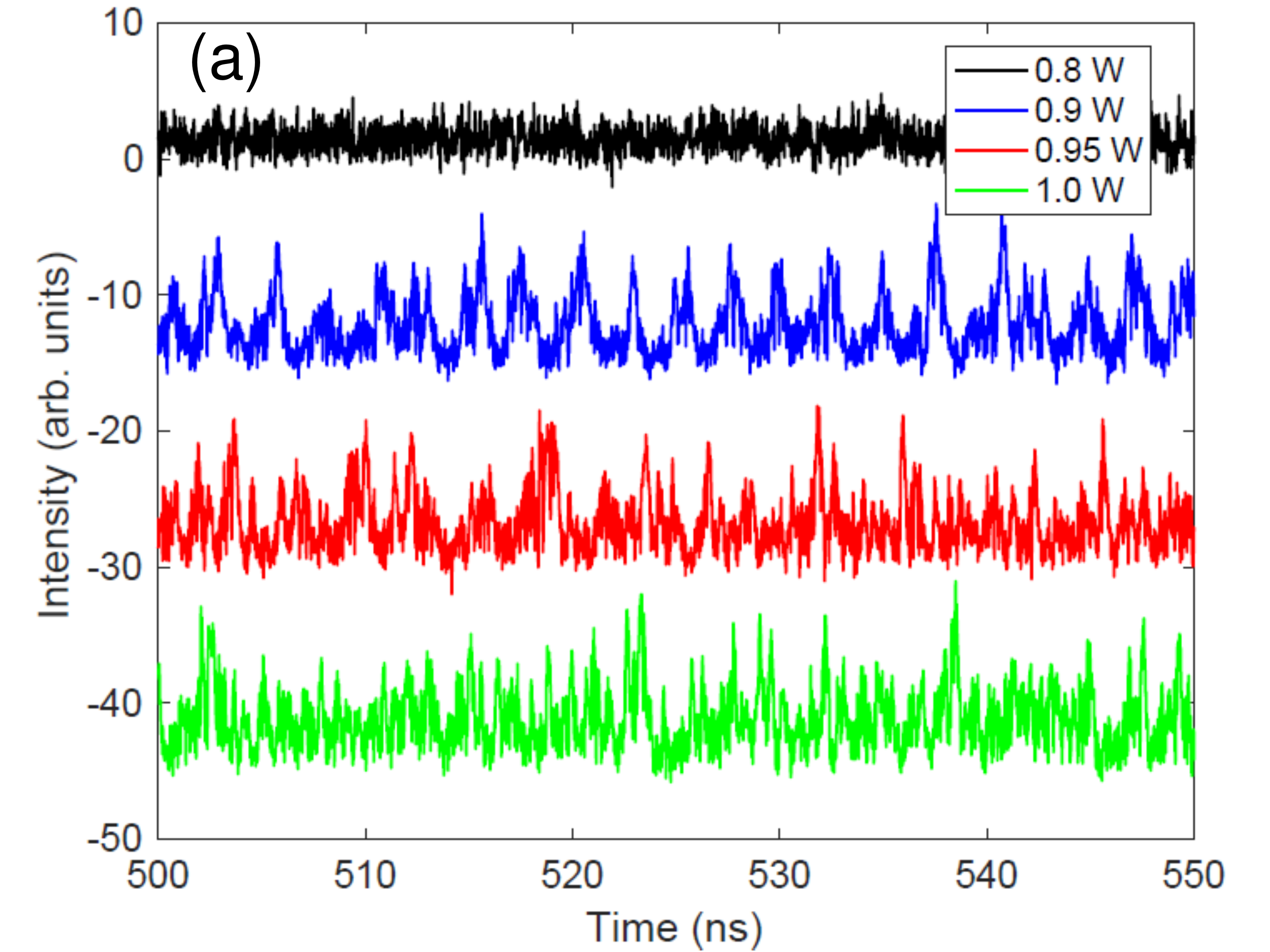}
\includegraphics[width=0.79\columnwidth]{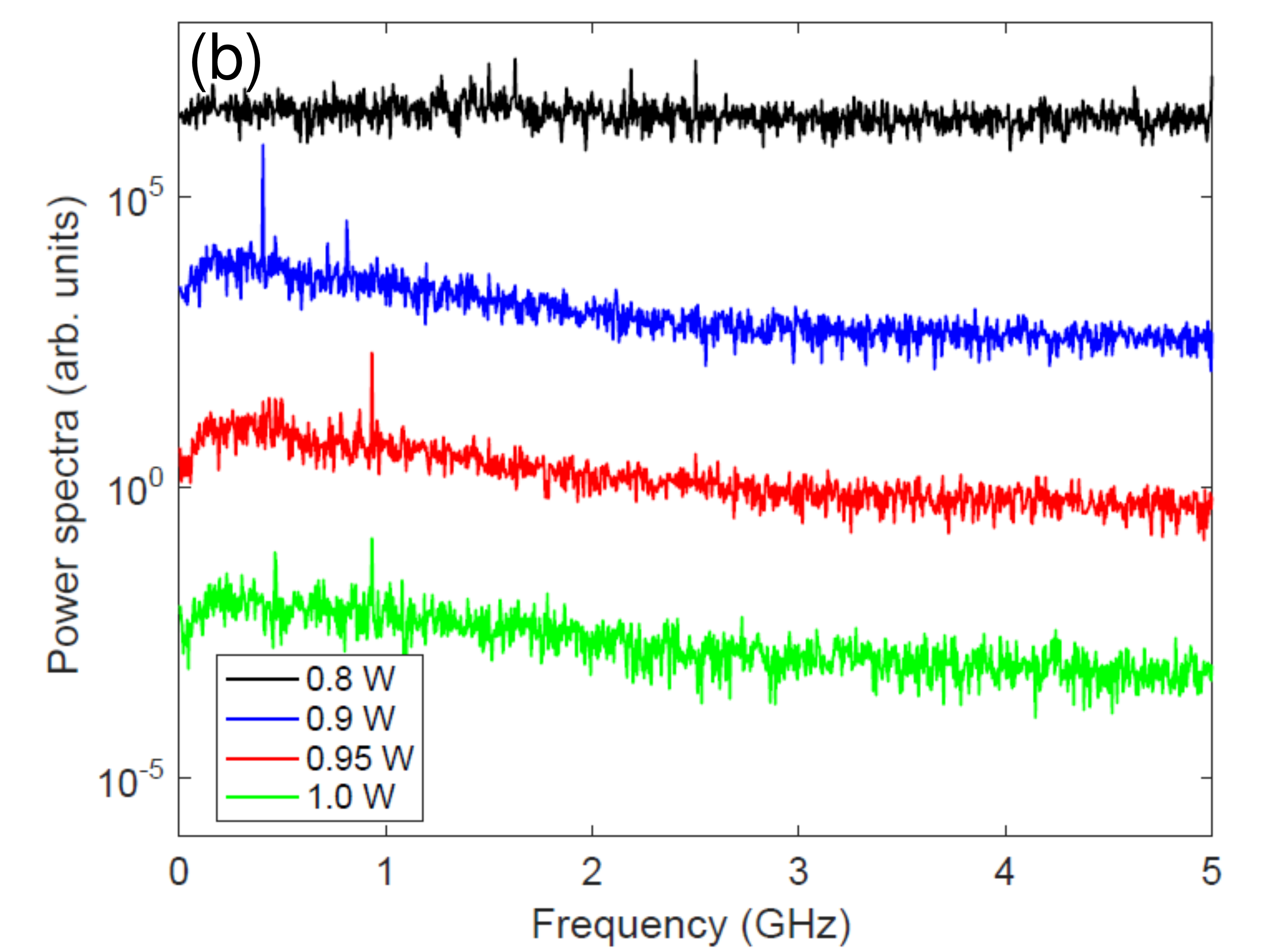}
\includegraphics[width=0.79\columnwidth]{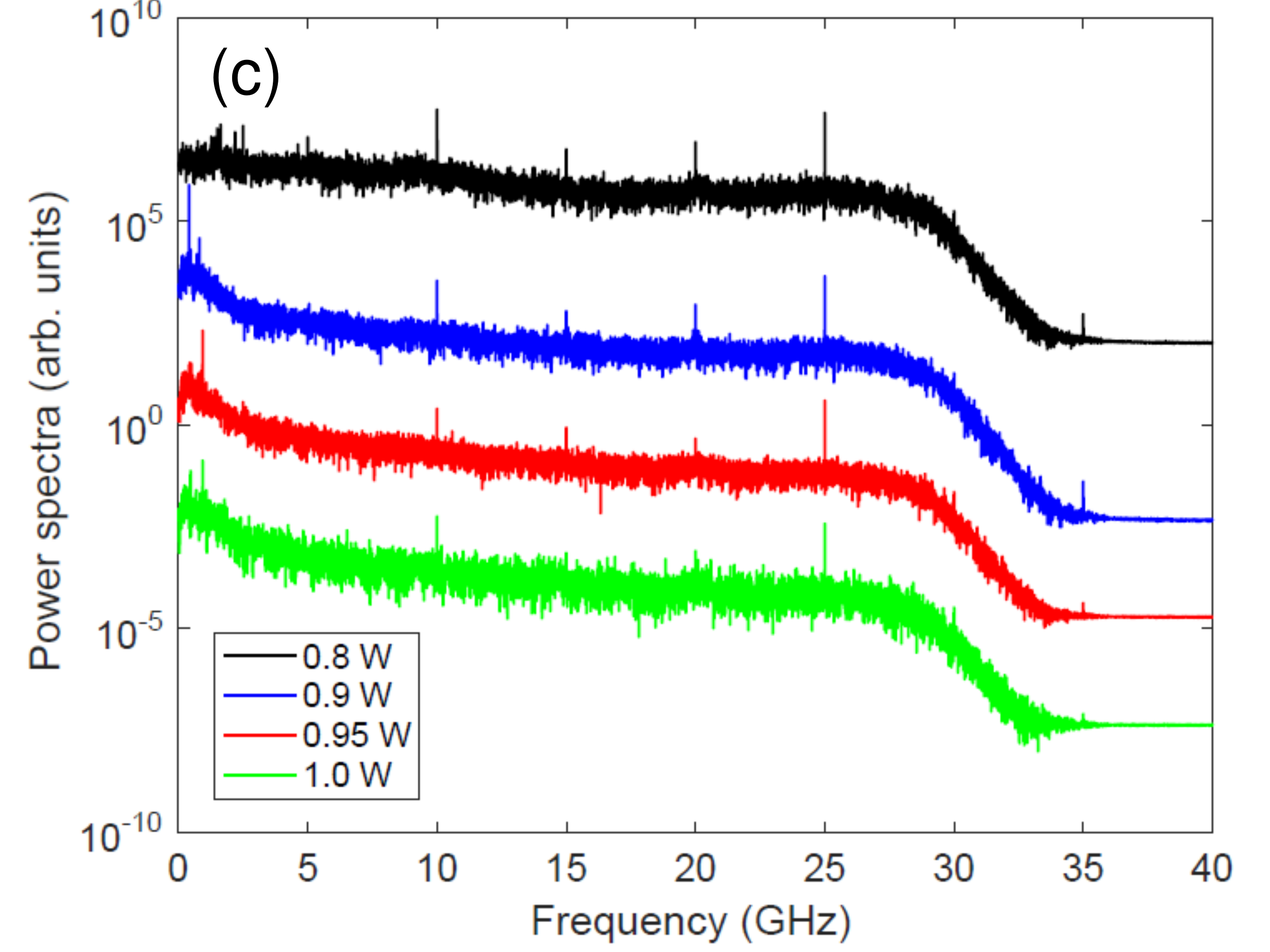}
\caption{(color online) (a) Intensity time series (separated vertically for clarity) below, during, and above the transition. (b) Power spectra at low frequencies (separated vertically for clarity), for the same pump powers as in panel (a). At the transition (0.9 W, second line, blue online) there is a narrow peak at $\sim$0.4 GHz, while for a slightly higher pump power (0.95 W, third line, red online) the peak is at $\sim$0.93 GHz. (c) As (b) but the horizontal axis covers all the range of frequencies.}\label{fig:ts_spectra}
\end{figure}

Examples of intensity time traces and the corresponding Fourier spectra are shown in Fig. \ref{fig:ts_spectra}. At the transition noisy oscillations are seen with a periodicity of about 2.5 ns, while in the broadband power spectrum there is a narrow peak at 0.4 GHz. As it will be shown latter, the appearance of this noisy periodicity can be understood in terms of a surprisingly simple model.

\section{Results}
\subsection{Horizontal Visibility Graph analysis}

First, we compare the empirical intensity time series with synthetic series generated by fBm, which is a family of processes, ${\textbf{B}^{\mathcal{H}}(t)}$, that is Gaussian, self-similar and endowed with stationary increments, $\textbf{W}^{\mathcal{H}}(t)=\textbf{B}^{\mathcal{H}}(t+1)-\textbf{B}^{\mathcal{H}}(t)$ (which are known as \textit{fractional Gaussian noise}, fGn). fBm has tunable memory and ${\textbf{B}^{\mathcal{H}=0.5}}$ corresponds to ordinary, memory-less Brownian motion, for which successive increments, $\textbf{W}^{\mathcal{H}=0.5}$, represent Gaussian white noise.

Using the HVG method (described in \textit{Appendix A}) we transform the empirical and synthetic time series into graphs, and then compare the graphs by comparing their degree distributions, $P(k)$. The simplest way to do this is to fit $P(k)$ to an exponential, $P(k)\propto \exp(-\lambda k)$, and compare the values of $\lambda$. A second way is based in comparing information measures computed from $P(k)$: the HVG entropy and the Fisher information \cite{ravetti}.

\begin{figure}
\centering
\includegraphics[width=0.8\columnwidth]{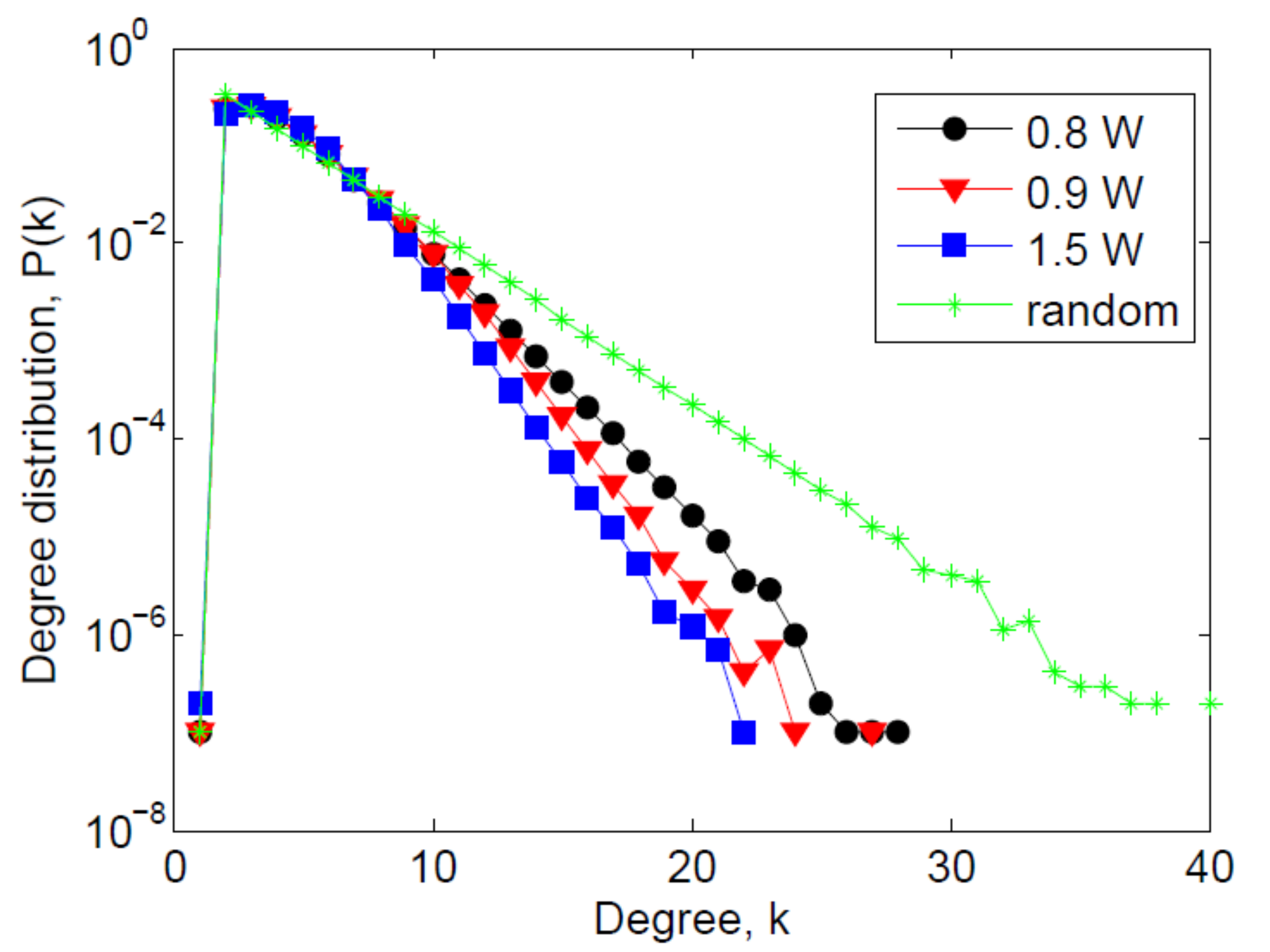}
\caption{HVG degree distributions of ``raw'' intensity time series for various pump powers. For comparison the distribution obtained from Gaussian white noise is also displayed.}\label{fig:hvg_degree}
\end{figure}

Figure~\ref{fig:hvg_degree}(a) displays the degree distribution, $P(k)$, obtained from the raw intensity time series below, at, and above the transition to turbulence, and also from Gaussian white noise. These distributions can be fitted to $\exp(-\lambda k)$ with $\lambda$ = 0.59, 0.69 and 0.75 for pump power 0.8~W (before), 0.9~W (at) and 1.5~W (after the transition). By comparing these $\lambda$ values with those obtained from fBm generated with different $\mathcal{H}$ values~\cite{ravetti}, and considering the same scaling region ($3 \leq k \leq 20$), we infer the Hurst exponent to be $\mathcal{H}$~=~0.3, 0.5 and $~0.6-0.7$ for 0.8~W, 0.9~W and 1.5~W, respectively. Therefore, the dynamics changes from anti-persistent (before the transition) to persistent (after the transition). At the transition, $\lambda$ is comparable to the one found for fBm with $\mathcal{H}=0.5$ that corresponds to pure Brownian motion.

In contrast, if we first threshold the raw data and keep only the intensity peaks that are higher than a certain threshold (as in~\cite{prl_2016}, we use a threshold equal to the intensity mean value plus two standard deviations) we find that the thresholded data has very different properties: $\lambda$ values are consistent with those found for fGn~\cite{ravetti} with $\lambda=0.403$ ($\mathcal{H}=0.5$) before the transition, $\lambda=0.43$ ($\mathcal{H}=0.7$) at the transition and $\lambda=0.395$ ($\mathcal{H}\sim 0.4-0.5$) after the transition.  $\lambda=0.43$ at the transition suggests a persistent fGn, while all other $\lambda$ values are close to the $\lambda$ expected for Gaussian white noise ($\lambda=0.405$). 

This effect of the threshold resembles the threshold sensitivity found in certain chaotic systems, where varying the threshold can lead from clustering to repelling of extreme events, or vice versa \cite{davidsen}.

The fit of $P(k)$ to an exponential has the drawback that detailed information about the shape of $P(k)$ is lost, and also, that $\lambda$ depends on the scaling region.  Therefore, to compare the distributions, $P(k)$, derived from empirical and from synthetic data we use information-theory measures: the HVG entropy, $S[P]$, and the Fisher information, $F[P]$ \cite{expressions}. In this approach, each time series is represented as a dot in the $S\times F$ plane~\cite{sfplane} and ``trajectories'' are obtained when the experimental parameter (laser pump power) and the synthetic parameter (Hurst exponent $\mathcal{H}$) are varied.

\begin{figure}
\centering
\includegraphics[width=0.8\columnwidth]{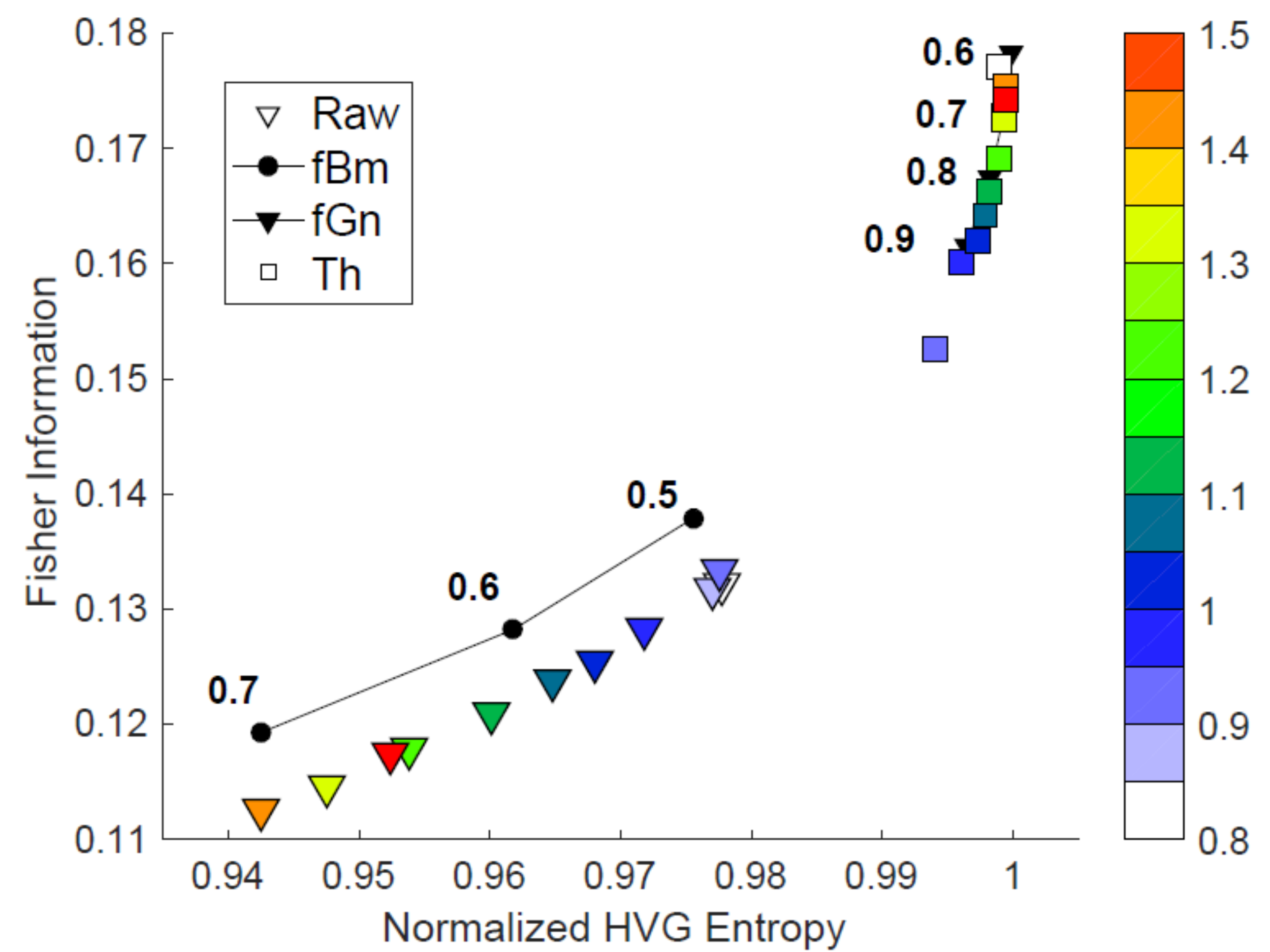}
\caption{The HVG entropy, $S$ (normalized to the entropy of Gaussian white noise), and the Fisher information, $F$, computed from the empirical intensity time series are compared with synthetic data. For the empirical data, the color code indicates the laser pump power in Watts; for the synthetic data (black symbols) the numbers indicate the Hurst exponent.}\label{fig:sfmap}
\end{figure}

In Fig.~\ref{fig:sfmap} the results of the analysis of the raw and the thresholded intensity data are compared to synthetic data (fBm and fGn). In the raw data there is a good agreement to what was inferred with the $\lambda$ fit: the raw data is well modeled by the fBm process; before and at the transition is close to fBm with $\mathcal{H}\sim 0.5$ (ordinary Brownian motion), while after the transition is close to fBm with $\mathcal{H}>0.5$, which indicates a persistent process. The thresholded intensity peaks are well represented by the fGn processes, also in good agreement with the $\lambda$ fit. The transition point (0.9 W) has the lowest $S$ and the lowest $F$ (is a return point of the trajectory in the $S\times F$ plane). The $\mathcal{H}$ values estimated from the comparison of the empirical and synthetic datasets in the $S\times F$ plane, and from the fit of $P(k)$ are quantitatively different, but agree qualitatively: for the raw data, both methods reveal a gradual transition, as the pump power increases, to a persistent process, while for the thresholded data, according to the $\lambda$ fit, at the transition $\mathcal{H}\sim 0.7$ while the $S\times F$ plane suggests $\mathcal{H}>0.9$.
 
In Fig.~\ref{fig:sfmap} we note that the raw intensity datasets are located close to the fBm ``trajectory'', but there is a more or less constant distance to this line, which indicates that fBm does not fully represent the empirical data. This is consistent with the analysis of \cite{BS_2007}, where the probabilities of ordinal patterns of length three were computed analytically for a fBm process, and it was shown that they have a particular symmetry [$P(012)=P(210)=p$, while the other four probabilities are equal to $(1/2-p)/2$], which does not hold for the probabilities computed from the empirical data [as it can be seen in~\cite{prl_2016}, Fig.~2(a)].

\subsection{Ordinal analysis}

Next, we show that the correlations uncovered in~\cite{prl_2016} by applying lagged ordinal analysis to the raw intensity data can be understood in terms of a surprisingly simple model. The ordinal method (described in \textit{Appendix A}) is used to uncover order relations among three intensity values $(I_i, I_{i+\tau}, I_{i+2\tau})$, where $\tau$ is an integer that gives an effective sampling time of $\tau dt$. In this way, each symbol encodes information about the intensity evolution during an interval of $3\tau dt$.

\begin{figure}
\centering
\includegraphics[width=1.0\columnwidth]{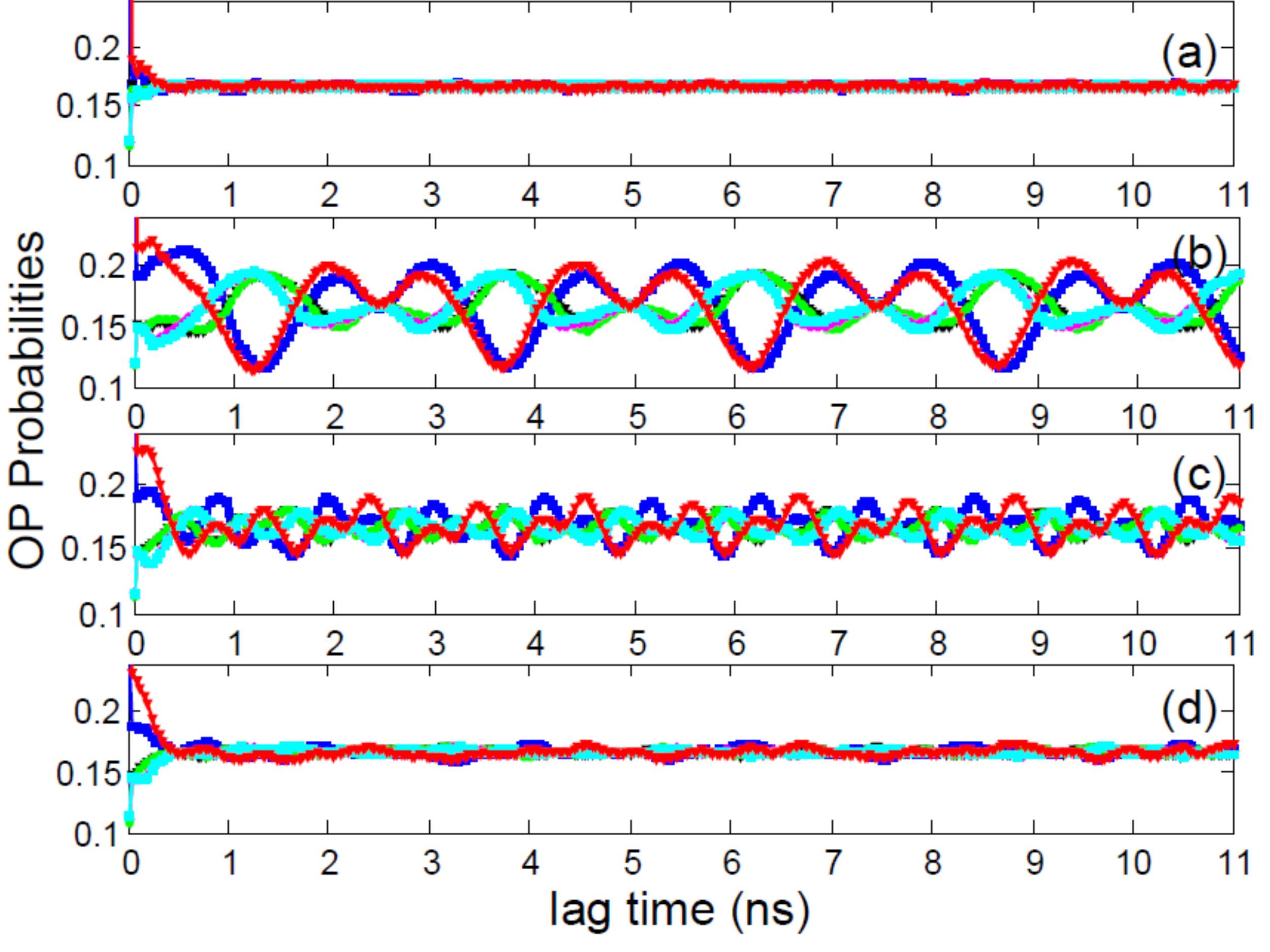}
\caption{Probabilities of the ordinal patterns computed from the laser intensity vs. the sampling time. The pump power is as in Fig. 1: (a) 0.8 W, (b) 0.9 W, (c) 0.95 W and (d) 1.0 W.}\label{fig:op_probs}
\end{figure}

The analysis of the ordinal probabilities vs. $\tau$, shown in Fig.~\ref{fig:op_probs}, reveals that below and above the transition there are no long-range correlations, as in panels \ref{fig:op_probs}(a) and \ref{fig:op_probs}(d), for $\tau$ large enough, the six patterns are equally probable. In contrast, at the transition [Fig.~\ref{fig:op_probs}(b)] the pattern probabilities oscillate regularly with periodicity of about 2.5 ns. For a pump power slightly above the transition, Fig.~\ref{fig:op_probs}(c), there are also regular oscillations of the probabilities with $\tau$, but the oscillations are of smaller amplitude.

Let us next show that these oscillations are captured by a remarkably simple model: a phase equation describing an stochastic limit cycle. 

Considering that the polar coordinates of a particle moving in a limit cycle trajectory are $a(t)e^{i\phi t}$ and neglecting the amplitude variations, the dynamics is described by a single rate equation for the phase:
\begin{equation}
d\phi/dt = \omega_0 + f(\phi,t)+\zeta(t),
\end{equation}
\noindent where $\omega_0$ is the angular rotation frequency, $f(\phi,t)$ is a $2\pi$ periodic function [$f(\phi,t)=f(\phi+2\pi,t)$] that represents the variability of the instantaneous frequency, and $\zeta$ represents stochastic fluctuations. 

By stroboscopic sampling every time interval $\Delta T$, the limit cycle evolution is described by a circle map \cite{libro_pik}:   
\begin{equation}
\phi(t+\Delta T) = \phi(t)+\omega_0\Delta T + F(\phi,t) +\xi(t),
\end{equation}
\noindent where $F(\phi,t)$ is a $2\pi$ periodic function that represents the phase accumulated over the time interval $\Delta T$, due to the variability of the instantaneous frequency, and $\xi$ represents the influence of the stochastic term. Assuming $F(\phi,t)=K\sin(\phi)$ gives  
\begin{equation}
\phi_{i+1} = \phi_i + \epsilon\rho + (K/2\pi)\sin(2\pi\phi_i) +D\xi_i.
\label{eqn:cm}
\end{equation}
\noindent where $\phi_i = \phi(t)$, $\phi_{i+1}=\phi(t+\Delta T)$, $\rho=\nu_0\Delta T$ with $\nu_0=\omega_0/2\pi$. We also include a parameter $\epsilon=\pm 1$ that determines the direction of the rotation (anticlockwise or clockwise). In the following we assume that $\xi_i$ a Gaussian white noise and $D$ is the noise strength. 

Next, we apply ordinal analysis to phase increments, $\Delta \phi_i=\phi_{i} - \phi_{i-1}$, generated from iterations of this map. We keep constant the strength of the nonlinearity, $K$, and the strength of the noise, $D$, and vary $\rho$ as control parameter. We chose $\rho$ because it is proportional to the stroboscopic sampling time, $\Delta T$, which is analogous, in the experimental situation, to the effective sampling time of the laser intensity, $\tau dt$, used to define ordinal patterns from lagged intensity values.

For appropriated values of $K$ and $D$ we find that the circle map gives a set of ordinal probabilities that are in remarkable agreement with those computed from the laser data at the transition.

\begin{figure}
\centering
\includegraphics[width=1.0\columnwidth]{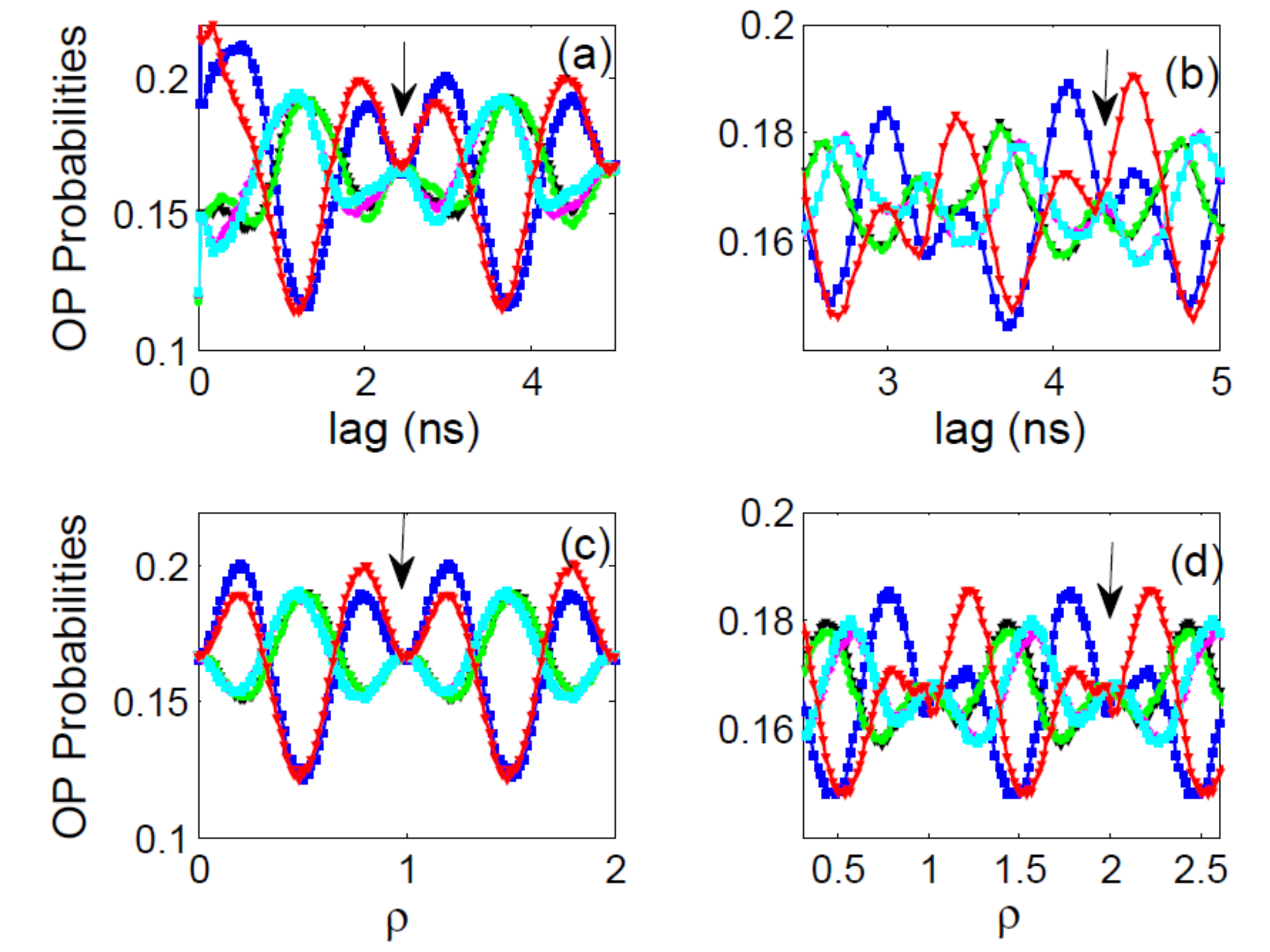}
\caption{Comparison of the ordinal probabilities computed from empirical data, panels (a) and (b), and from synthetic data, panels (c) and (d). In  (a) and (b), the probabilities are computed from the laser intensity time-series and the horizontal axis is the sampling time; the laser operating condition is (a) at the transition and (b) slightly above the transition (pump power 0.9 W and 0.95 W, respectively). In panels (c) and (d), the probabilities are computed from data generated by iterating the circle map, Eq. (\ref{eqn:cm}), and the horizontal axis is the map parameter, $\rho$; the other parameters are (c) $K=0.1$, $D=0.02$, $\epsilon=1$; (d) $K=0.25$, $D=0.075$, $\epsilon=-1$.}\label{fig:comparison}
\end{figure}

Figure~\ref{fig:comparison} allows a precise comparison: panels (a) and (b) display in detail the oscillatory behavior of the probabilities with the effective sampling time of the intensity time series, while panels (c) and (d) display the probabilities computed from iterations of the circle map. We observe an excellent agreement as the same hierarchical structure (more/less probable patterns) and clustered structure (pairs of patterns with the same probability) are seen when comparing the empirical and the synthetic data. We note that the ordinal probabilities at the transition, Fig.~\ref{fig:comparison}(a), are reproduced by the iterations of the circle map with $\epsilon=1$, Fig.~\ref{fig:comparison}(c), while slightly above the transition, Fig. \ref{fig:comparison}(b), with $\epsilon=-1$, Fig. \ref{fig:comparison}(d). This suggests that immediately after the transition there is a change of rotation.

Contrasting similar situations in Figs.~\ref{fig:comparison}(a) and \ref{fig:comparison}(c), we observe that $\rho=1$ in the circle map data corresponds to $\tau=2.5$ ns in the laser data (as indicated with arrows). Because $\rho=\nu_0 \tau$, using $\rho=1$ and $\tau=2.5$ ns we can estimate the frequency of the rotation in the limit cycle as $\nu_0=1/\tau=0.4$ GHz, in agreement with the narrow peak seen in the spectrum in Fig.~\ref{fig:ts_spectra}. Also comparing similar situations immediately after the transition, in Figs.~\ref{fig:comparison}(b) and \ref{fig:comparison}(d), we observe that $\rho=2$ in the circle map data corresponds to $\tau=4.3$ ns in the laser data (as indicated with arrows). The same argument gives $\nu_0=2/\tau=0.46$ GHz, and the spectrum in Fig.~\ref{fig:ts_spectra} we see the peak at about 0.93~GHz, which is consistent with $2\nu_0$. 

The agreement found is unexpected because, as shown in Fig.~\ref{fig:ts_spectra}, the spectrum is extremely broad and thus, it is surprising to find that, in the symbolic representation, the intensity temporal dynamics is described by an stochastic limit cycle with rotation frequency $\nu_0$. It is worth noticing that statistics of the intensity values are not described by the statistics of the phase increments, $\Delta \phi_i=\phi_{i} - \phi_{i-1}$, which are positive or negative. 

\section{Conclusions}

To summarize, we have applied two data analysis tools to characterize persistence and temporal correlations in the intensity dynamics of a fiber laser. To characterize the persistence, intensity time series (raw and thresholded data) were transformed to graphs through the horizontal visibility algorithm and then compared with well-known stochastic processes, fractional Brownian motions (fBm) and fractional Gaussian noises (fGn). 

Two different techniques that use the graph degree distribution (fitting the distribution to an exponential, and computing, from the degree distribution, the Shannon entropy and the Fisher information) gave consistent results, with the raw intensity data being modeled by a fBm processes, and the thresholded data, by a fGn. The analysis of the raw data revealed that the dynamics is very close to the fBm process, but there is a distance that suggests a degree of determinism in the dynamics. With respect to the thresholded data, the HVG analysis reveals that the empirical data is fully consistent with an stochastic fGn process.

Using ordinal analysis we have also demonstrated that at the transition, correlations among three lagged intensity values can be precisely represented by a surprisingly simple model: a circle map. The physics underlying the emergence of stochastic periodicity at the transition could be mode locking; however, is computationally unfeasible to simulate the fiber laser model with a realistic number of modes (up to a million in the turbulent region). Therefore, in \textit{Appendix B}, in order to check the generality of our results, we analyze two multi-mode laser models and find, for certain parameters, a variation of the ordinal probabilities with the lag, similar to that found in the empirical laser data at the transition.

Because of the unavailability of a model that reproduces the fiber laser dynamics in a time scale shorter than the cavity round trip time, our work is limited to the characterization of the empirical data, and leaves several relevant open questions: the physical origin of the temporal correlations, the relation between the characteristic time scale and the physical laser parameters, and the reason why the statistics of the raw and thresholded time series are different. We hope that our results will motivate further experimental and theoretical studies to address these issues.

Extracting information from observed data is a main challenge in diverse areas of engineering and science, and the analysis tools used here can be very useful for investigating the output signals of other systems that undergo similar transitions to turbulent regimes \cite{prl_2016_2,nat_comm_2017}. They provide complementary insights into the correlations present in the data: the HVG method is parameter-free and captures correlations whose range is limited only by the actual values of the data points; on the other hand, the ordinal method (that only considers the temporal order of the data points but not their actual values) has two parameters (the length of the pattern and the lag between the data points) which allow tuning the scale of the analysis. 

\section{Acknowledgments}
We gratefully acknowledge Prof. S. K. Turitsyn (Aston University, U.K.) for the permission to use the datasets analysed in \cite{prl_2016}. L. C. acknowledges support from Brazilian agencies FAPEMIG (Project Number: APQ-03664-16), CNPq and CAPES. C. M. acknowledges partial support from Spanish MINECO (FIS2015-66503-C3-2-P) and from the program ICREA ACADEMIA of Generalitat de Catalunya.

\section{Appendix A: Methods of time-series analysis}

In this Appendix we describe the two methods used in the main text to uncover temporal correlations in the intensity dynamics: the Horizontal Visibility Graph \cite{Lacasa2009} and Ordinal Analysis \cite{bandt_PRL_2002}. 

\subsection{Horizontal Visibility Graph}

This method transforms a real time-series, $x=\{x_1, \dots x_i, \dots x_N\}$, into a graph, by connecting pairs of data points if there is ``horizontal visibility'' between them: $x_i$ and $x_j$ are connected if $x_i>x_k$ and $x_j>x_k$ for all $k$ such that $i<k<j$.

As an example, we consider the following time series:
\begin{eqnarray}
x = \{0.71, \mbox{  }0.53, \mbox{  }0.56, \mbox{  }0.89, \mbox{  }0.50, \nonumber \\
\mbox{  }0.77, \mbox{  }0.21, \mbox{  }0.6, \mbox{  }0.72, \mbox{  }0.35 \}.
\label{eq:time-series}
\end{eqnarray} 

The number of links, $k_i$, that each data point, $x_i$, has is
\begin{equation}
k = \{3,\mbox{ } 2,\mbox{ } 3,\mbox{ } 4,\mbox{ } 2,\mbox{ } 5,\mbox{ } 2,\mbox{ } 3,\mbox{ } 3,\mbox{ } 1 \}.
\end{equation} 

\begin{figure}[htb]
\centering
\includegraphics[width=0.6\columnwidth]{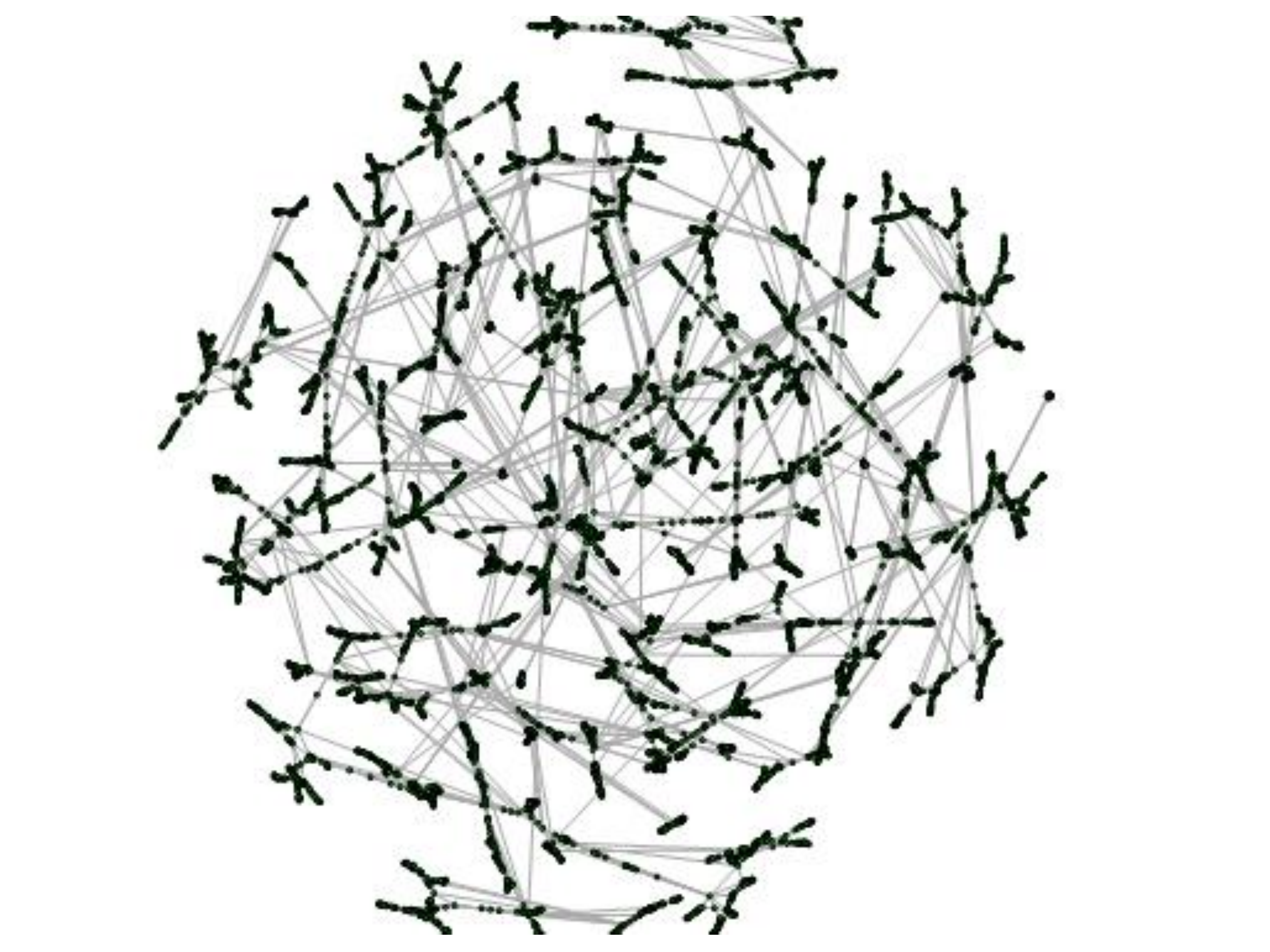}
\caption{\label{fig:hvg}Horizontal Visibility Graph obtained from empirical laser data.}
\end{figure}

The degree distribution, $P(k)$, is the probability that a data point has $k$ links. 
As an example, Fig.~\ref{fig:hvg} displays a graph that is obtained from the empirical laser data.

\subsection{Ordinal Analysis}

\begin{figure}[htb]
\centering
\includegraphics[width=0.9\columnwidth]{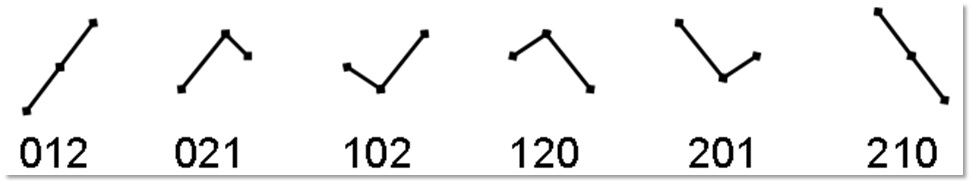}
\caption{The six possible ordinal patterns that can be obtained from $D=3$ data points.}
\label{fig:op}
\end{figure}

This method transforms a real time-series, $x=\{x_1, \dots x_i, \dots x_N\}$, into a sequence of symbols (known as \textit{ordinal patterns}), which are determined by considering the temporal order in which $D$ data points occur in the time-series. To fix the ideas, considering pairs of consecutive values, $x_i<x_{i+1}$ gives symbol `01', while $x_i>x_{i+1}$ gives `10'. If we consider $D=3$ values, as shown in Fig.~\ref{fig:op}, there are 6 possible symbols: $x_i<x_{i+1}<x_{i+2}$ gives `012', $x_i<x_{i+2}<x_{i+1}$ gives `021', etc. When two data values are identical, a small random value is added to one of them to break the symmetry.

More in general, $D!$ symbols represent the possible order relations among $D$ data points. By labeling the symbols from 1 to $D!$, a time series with $N$ real data values is transformed into a sequence of $N-D$ integers.

As an example, we consider the same time-series used in the previous section,
\begin{eqnarray}
x = \{0.71, \mbox{  }0.53, \mbox{  }0.56, \mbox{  }0.89, \mbox{  }0.50, \nonumber \\
\mbox{  }0.77, \mbox{  }0.21, \mbox{  }0.6, \mbox{  }0.72, \mbox{  }0.35 \}.
\end{eqnarray} 
The symbolic sequence obtained by considering the ordering of $D=3$ consecutive values is
\begin{equation}
s = \{201, \mbox{  }012, \mbox{  }120, \mbox{  }201, \mbox{  }120, \mbox{  }201, \mbox{  }012, \mbox{  }120 \},
\end{equation} 
from where we obtain the following set of probabilities (ordinal probabilities)
\begin{eqnarray}
P(012)&=&2/8, \mbox{  }P(021)=0, \mbox{  }P(102)=0, \nonumber\\
P(120)&=&3/8, \mbox{  }P(201)=3/8, \mbox{  }P(210)=0. \nonumber
\end{eqnarray}

By considering a lag the ordinal method can be used to analyze order relations among non-consecutive data values. In this case the ordinal patterns are defined from the temporal ordering of $D$ lagged data points, $x_i, x_{i+\tau}, \dots, x_{i+D\tau}$, where $\tau$ is an integer that, if $dt$ is the sampling time, gives an effective sampling time of $\tau dt$.

In order to gain insight into the probabilities computed when data points are not consecutive but have a lag between them, we analyze the case of a noisy periodic signal: we consider a synthetic time-series generated from a sinusoidal of period $T=1$,
\begin{equation}
x (t) = \sin (2\pi t) + \eta\xi,
\end{equation} 
where $\xi$ is a Gaussian white noise with zero-mean and unit variance, and $\eta$ is the noise strength. 

\begin{figure}[htb]
\centering
\includegraphics[width=0.9\columnwidth]{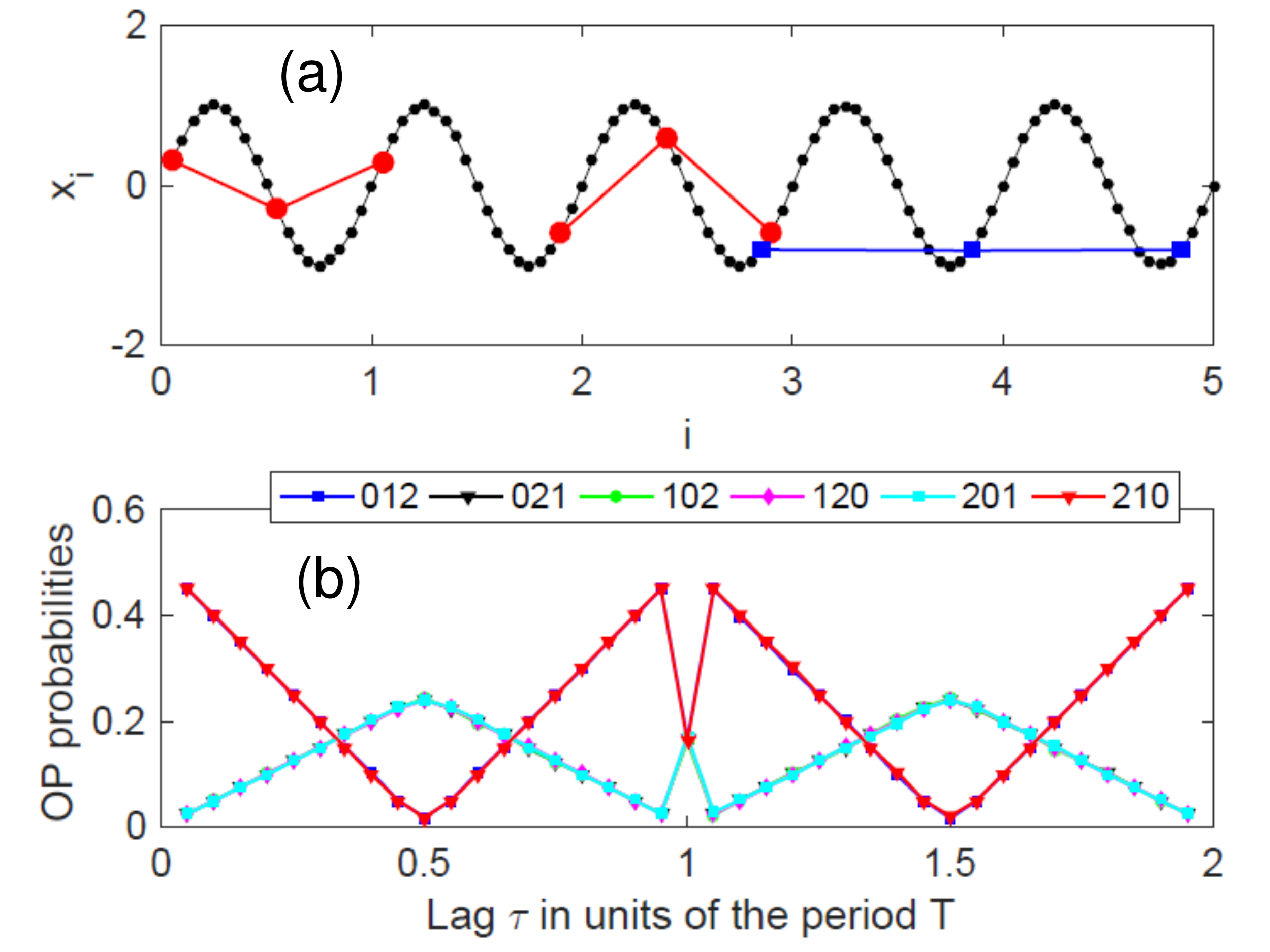}
\caption{Top: synthetic time series generated by sampling a sinusoidal of unit amplitude and period $T=1$, which has a small stochastic term added (the noise strength $\eta=0.01$). Three data points (filled circles) lagged by $\tau=T/2$ give either a ``V'' pattern (102, 201) or  a ``lambda'' pattern (120, 021). The filled squares indicate three data points lagged by $\tau=T$, which give, due to the small added noise, any pattern with equal probability. Bottom: ordinal probabilities computed from the synthetic time series, vs. the lag. We note that for $\tau=T/2$ and $\tau=3T/2$ the ``V'' and ``lambda'' patterns have high probability, while for $\tau=T$ all patterns are equally probable.}
\label{fig:opexp}
\end{figure}

We first generate a time-series of $N=4000$ data points with $\eta=0.01$ (small noise in comparison with the oscillation amplitude, which is equal to 1). The data points are sampled with $\Delta t= 1/20$ (i.e., the time-series covers 200 periods). Figure~\ref{fig:opexp} displays a short section of the time-series (five periods) and the probabilities of the six $D=3$ ordinal patterns vs. the lag (the integer numbers between 1 and 6 label the six patterns shown in Fig.~\ref{fig:op}). We note that the probabilities are divided in two groups: in one group are the ``V'' patterns (102, 201) and the ``lambda'' patterns (120, 021), while the other group includes the two ``trend'' patterns (012 and 210). The groups are either more or less probable, depending on the lag. An inspection of the time-series allows to gain insight into this dependency: if three data points are lagged by $\tau=T/2$ they form either a ``V'' or a ``lambda'' pattern, while if the three data points are lagged by $\tau=T$, they are equal if $\eta=0$ (as we have a pure sinusoidal of period $T$), but when $\eta\ne0$ the differences among the three data points are fully random and thus, their temporal order is random. Therefore, when $\tau=T$ all patterns are equally probable. 

\begin{figure}[htb]
\centering
\includegraphics[width=0.9\columnwidth]{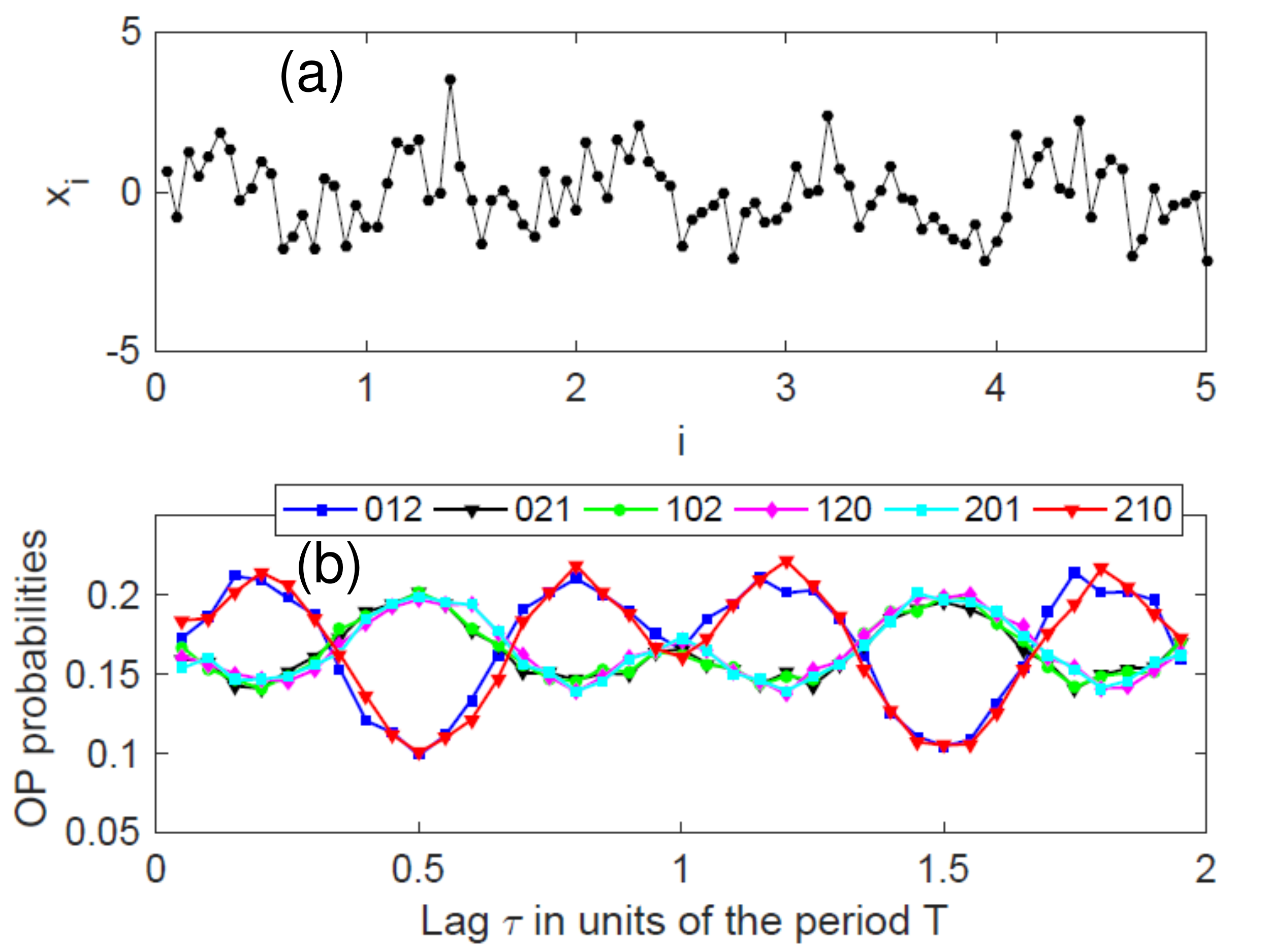}
\caption{As Fig.~\ref{fig:opexp} but when stronger noise, $\eta=1$, is added to the sinusoidal.}
\label{fig:opexp2}
\end{figure}

Next, we consider stronger noise. As seen in Fig.~\ref{fig:opexp2}, which is done in the same way but with $\eta=1$, the variation of the ordinal probabilities with the lag is more smooth, and resembles that found in the empirical laser data in the main text.

\section{Appendix B: Multi-mode laser models}

In this Appendix we present simulations of different laser models, and compare the results with those obtained from the empirical laser data (presented in the main text).

We performed extensive simulations of the laser model used in \cite{natphot}, however, with a limited number of modes this model only gives a good description of the temporal dynamics over many round-trips (considering the space-time representation used in \cite{natphot}, the model describes the dynamics in the temporal-like dimension), but fails to describe the fast temporal evolution (in the space-like dimension), and we were unable to reproduce the correlations studied in the main text.
 
As is computationally unfeasible to simulate the model with a realistic number of modes (up to a million in the turbulent region), we have used other laser models to check the generality of our results.

We first use a multi-longitudinal-mode model where modal competition leads to an oscillatory behavior of the total intensity \cite{mandel}, and then we use a two-mode model, which is well-known to describe the dynamics of two orthogonal linear polarization modes of vertical cavity surface emitting lasers (VCSELs) \cite{martin}. In both cases we show that, for appropriated parameters, the models predict an output intensity that has a similar variation of the ordinal probabilities with the lag, as that described for the fiber laser in the main text. Understanding the underlying mechanisms will require detailed investigations, as they will likely be different in the different models. However, our goal here is only to demonstrate that the correlations detected with ordinal analysis in the empirical data (fiber laser), can also be found in the output of two multimode laser models. A detailed analysis of the models' predictions is in progress and will be reported elsewhere.

\subsection{Multi-longitudinal-mode model}

In Ref.~\cite{mandel} mode switching in a multimode semiconductor laser was studied experimentally and it was observed that the intensity of each mode displays large amplitude oscillations, which obey a highly organized antiphase dynamics leading to an almost constant total intensity output. A multimode model was proposed that identified four-wave mixing as the dominant mechanism at the origin of the observed dynamics. Here we simulate the model rate equations that couple $N$ modal complex fields, $A_m(t)$, with $N$ spatial harmonics of the carrier density, $F_m(t)\sim \int N(z,t) \phi_m^2(z) dz$, that represent the grating in the carrier density created by the standing waves of the field. The model equations are

\begin{eqnarray}
\frac{dA_m}{dt}=(G_m-1)A_m - i\sigma  \sum_{k,p} \frac{A_k A_p A^*_{k+p-m}}{\eta(p-m)}, \nonumber\\
\eta\frac{dF_m}{dt}= J - F_m\left(1+\sum_n \beta_{mn}|A_n|^2\right).\nonumber
\end{eqnarray}

Here $G_m=F_m(1-\epsilon \sum_n \tilde{\beta}_{mn}|A_n|^2)$, $\tilde{\beta}_{mn}$ and $\beta_{mn}$ are self- and cross-coupling coefficients, $\tilde{\beta}_{mm}=\beta_{mm}=1$, $\tilde{\beta}_{mn}=(4/3)\beta_{mn}$. $\sigma$ is the strength of the four-wave-mixing term; the restriction in the double sum is that $1 \le k+p-m \le N$. Time is in units of the cavity round trip time, $\tau_p$, and $\eta$ is the ratio between $\tau_p$ and the carrier life time, $\tau_N$. $J$ is the pump normalized to the threshold value. The model in Ref.~\cite{mandel} also includes the well known $\alpha$ factor that accounts for symmetry breaking leading to a modal switching sequence from the blue to the red side of the optical spectrum. Here for simplicity we take $\alpha=0$.

The model equations were simulated with the same parameters as in \cite{mandel}: $J=1.5$, $\eta=1000$, $\tau_p=0.01$~ns, $\epsilon=0.05$, $\beta_{mn}=0.975$ ($\forall$ $m$, $n$, $m\ne n$), and $\sigma = 0.35$. Results for 4 modes are presented in Fig. 1. The variation of the ordinal probabilities with the lag is qualitatively very similar to that shown in the main text. Similar results are found with a higher number of modes, as shown in Fig.~\ref{fig:mandel_16}. Moreover, the plot of the entropy ($H=-\sum p_i \log p_i$) computed from the ordinal probabilities (known as \textit{permutation entropy} \cite{bandt_PRL_2002}) as a function of the lag and of the number of modes, shown in Fig.~\ref{fig:entropy}, suggests that the length of the correlations is maximum for a particular number of modes (for the parameters in Fig.~\ref{fig:entropy}, with 8 modes the entropy displays the longest oscillatory behavior with the lag). A detailed characterization of this effect is out of the scope of the present work and is left for future work.

\begin{figure*}[htb]
\centering
\includegraphics[width=0.67\columnwidth]{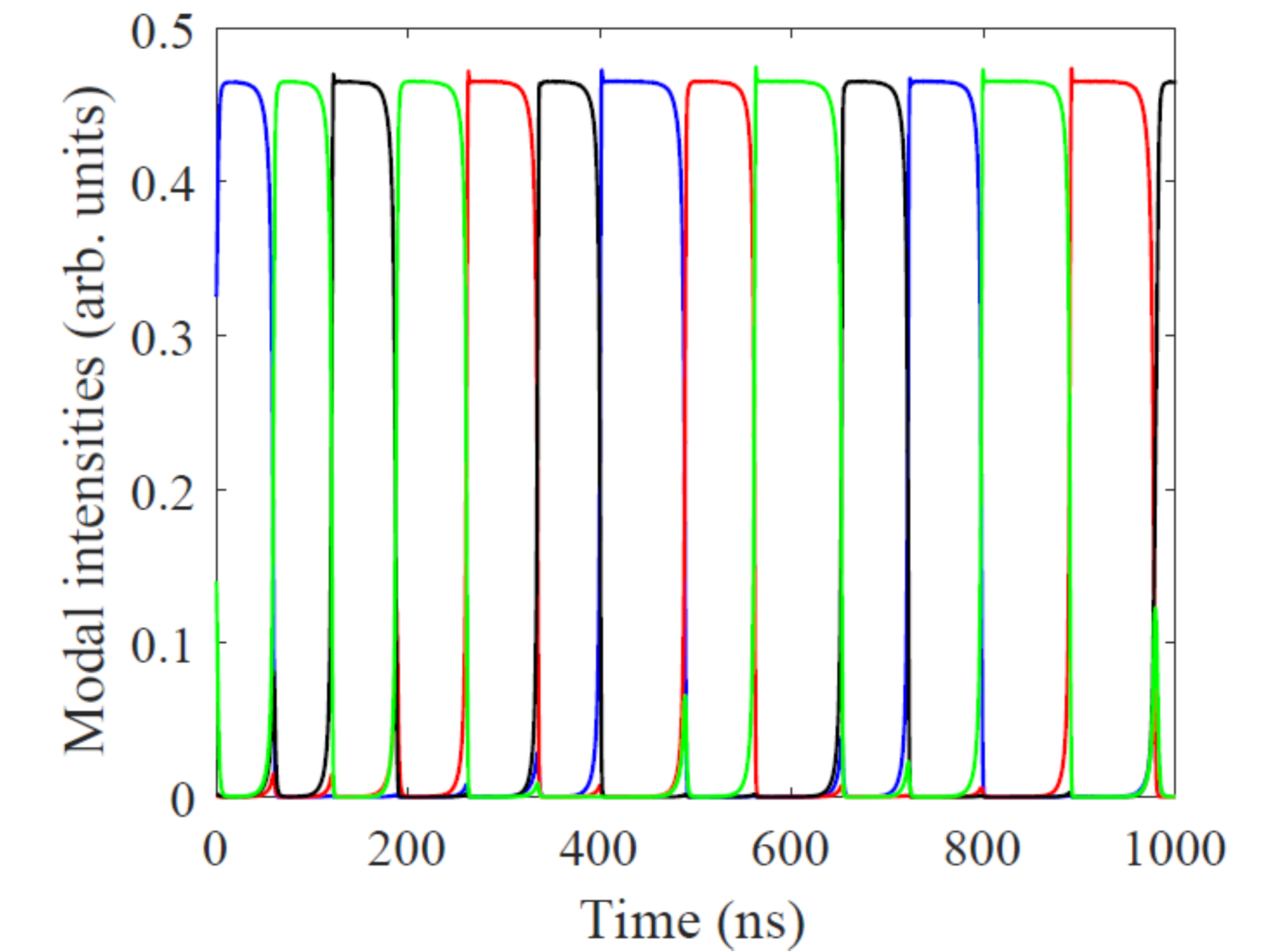}
\includegraphics[width=0.67\columnwidth]{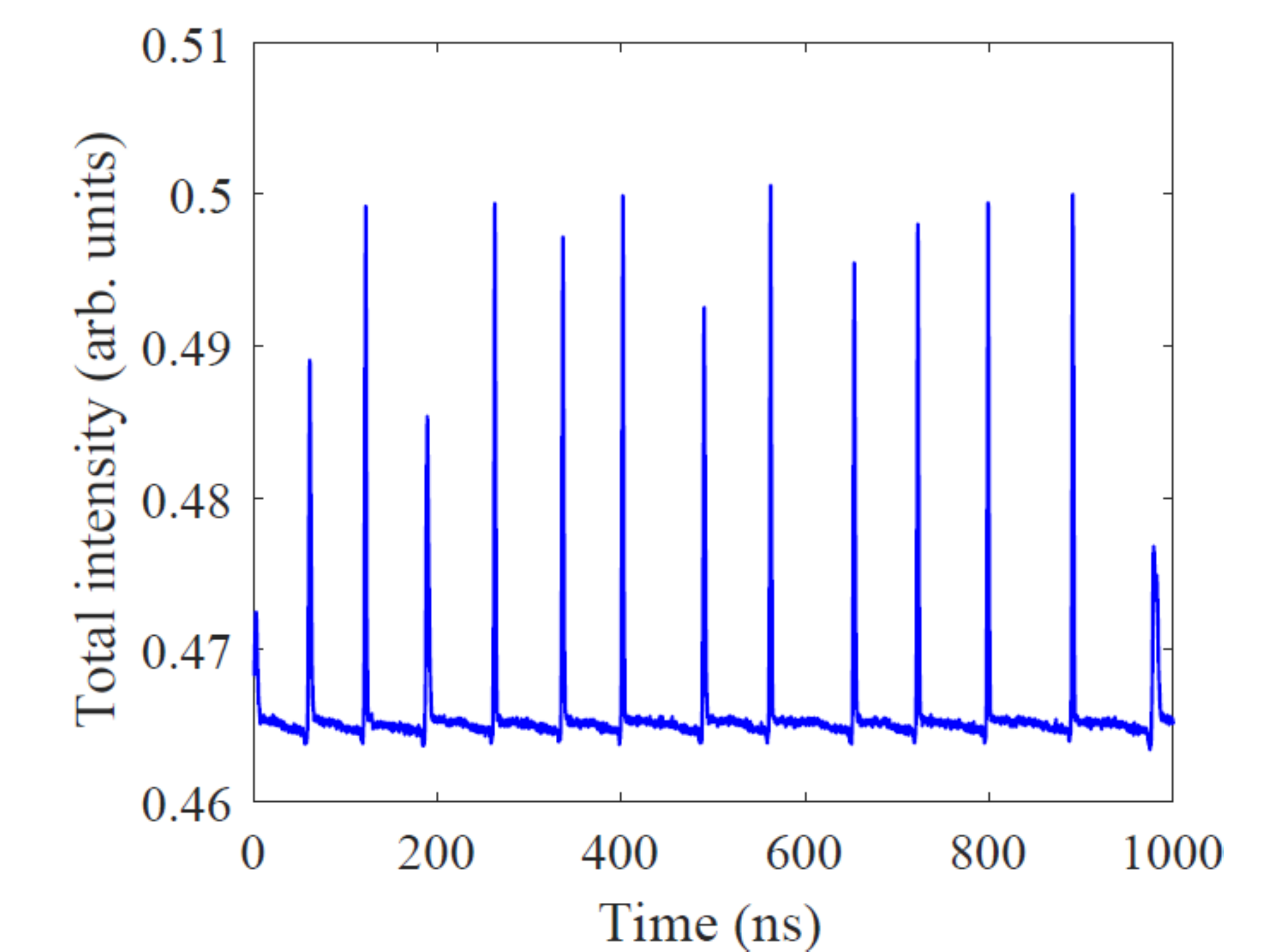}
\includegraphics[width=0.67\columnwidth]{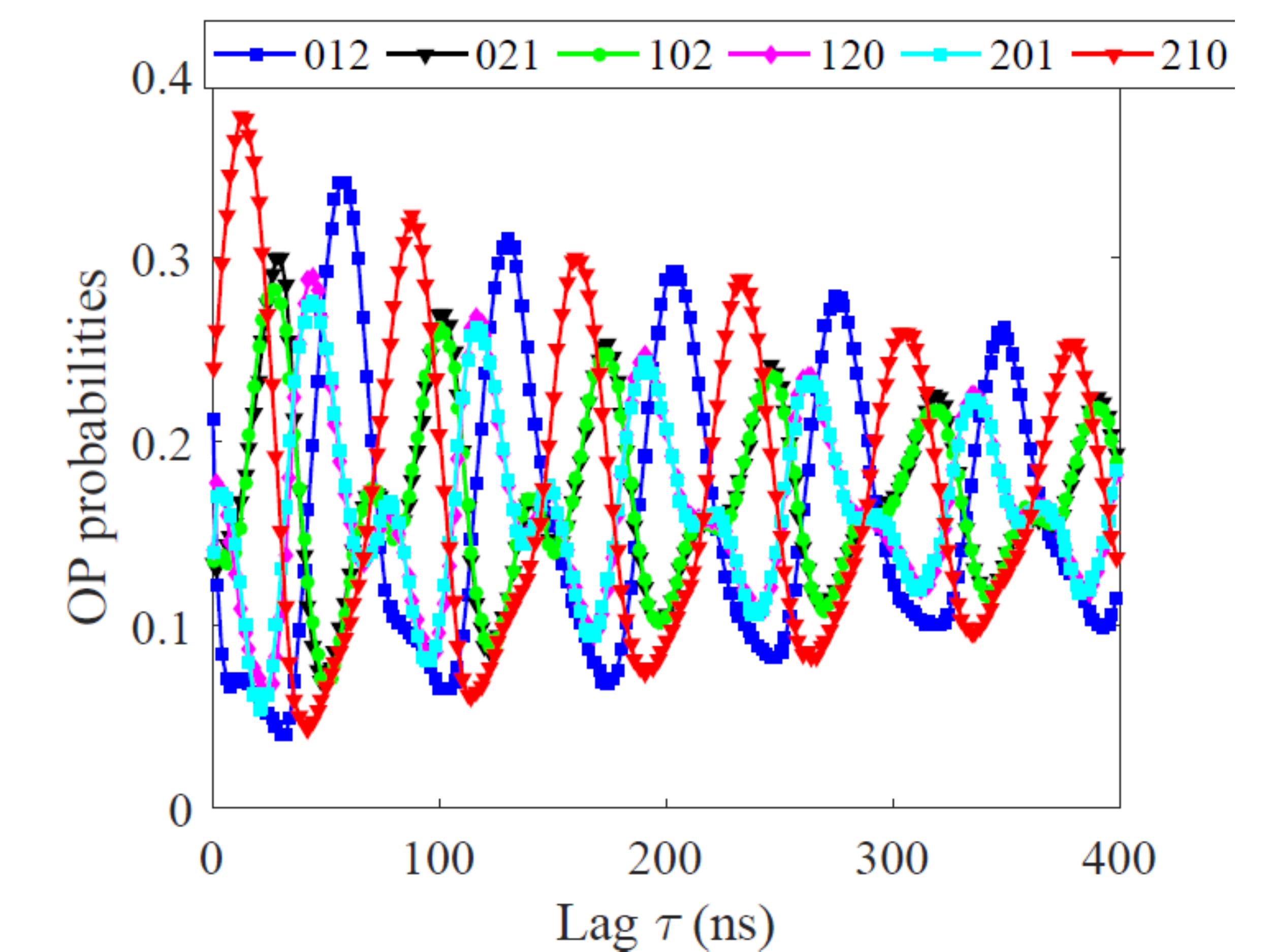}
\caption{Numerical results obtained from simulation of the multi-mode model with $N=4$ modes. The modal intensities (top) and the total intensity (center) are plotted vs. time. The bottom panel displays the ordinal probabilities (computed from the total intensity time-series, with ordinal patterns determined by the ordering of 3 intensity data points separated a lag $\tau$) vs. the lag $\tau$.}\label{fig:mandel_4}
\end{figure*}
\begin{figure}[htb]
\centering
\includegraphics[width=0.49\columnwidth]{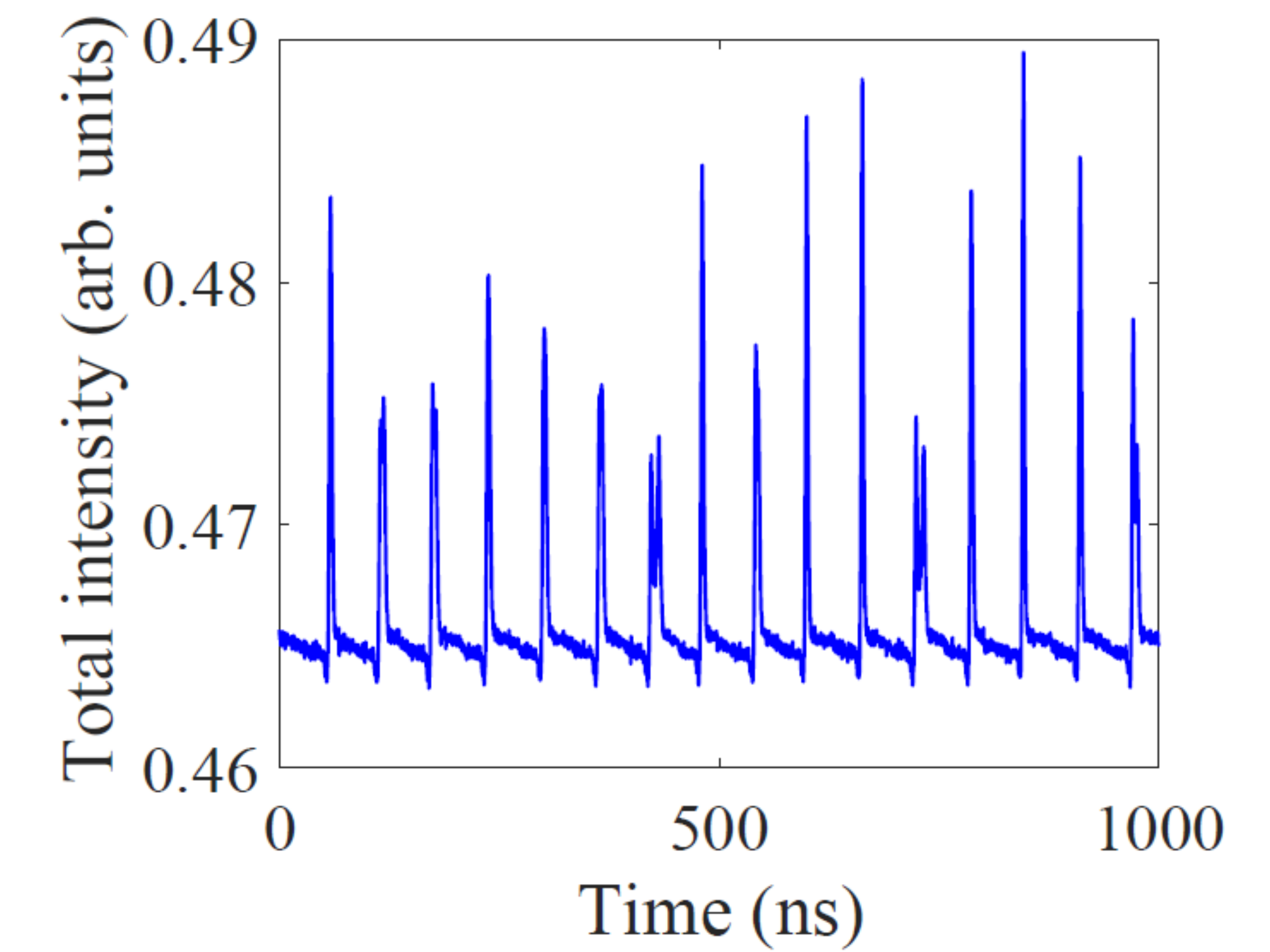}
\includegraphics[width=0.49\columnwidth]{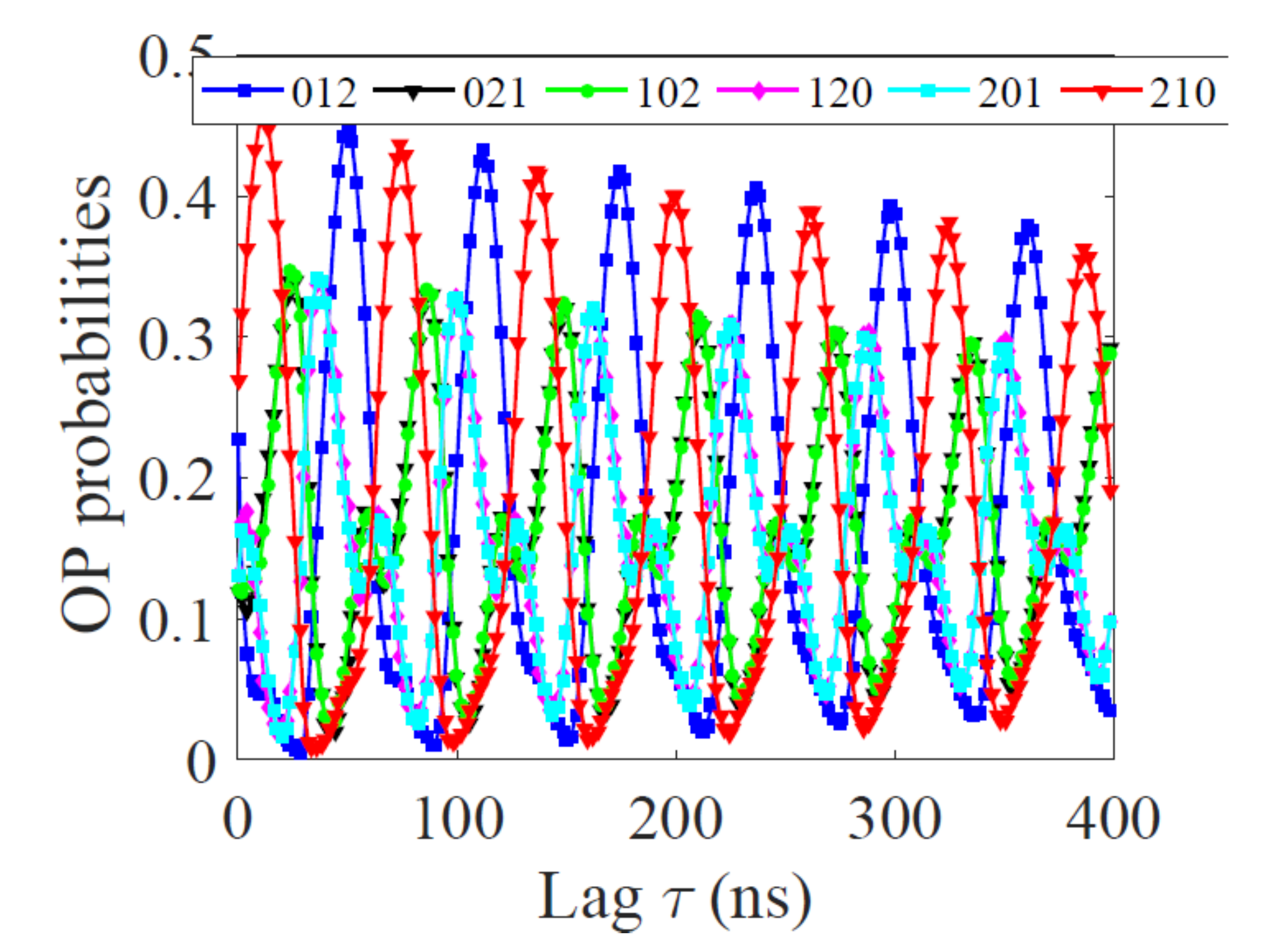}\\
\includegraphics[width=0.49\columnwidth]{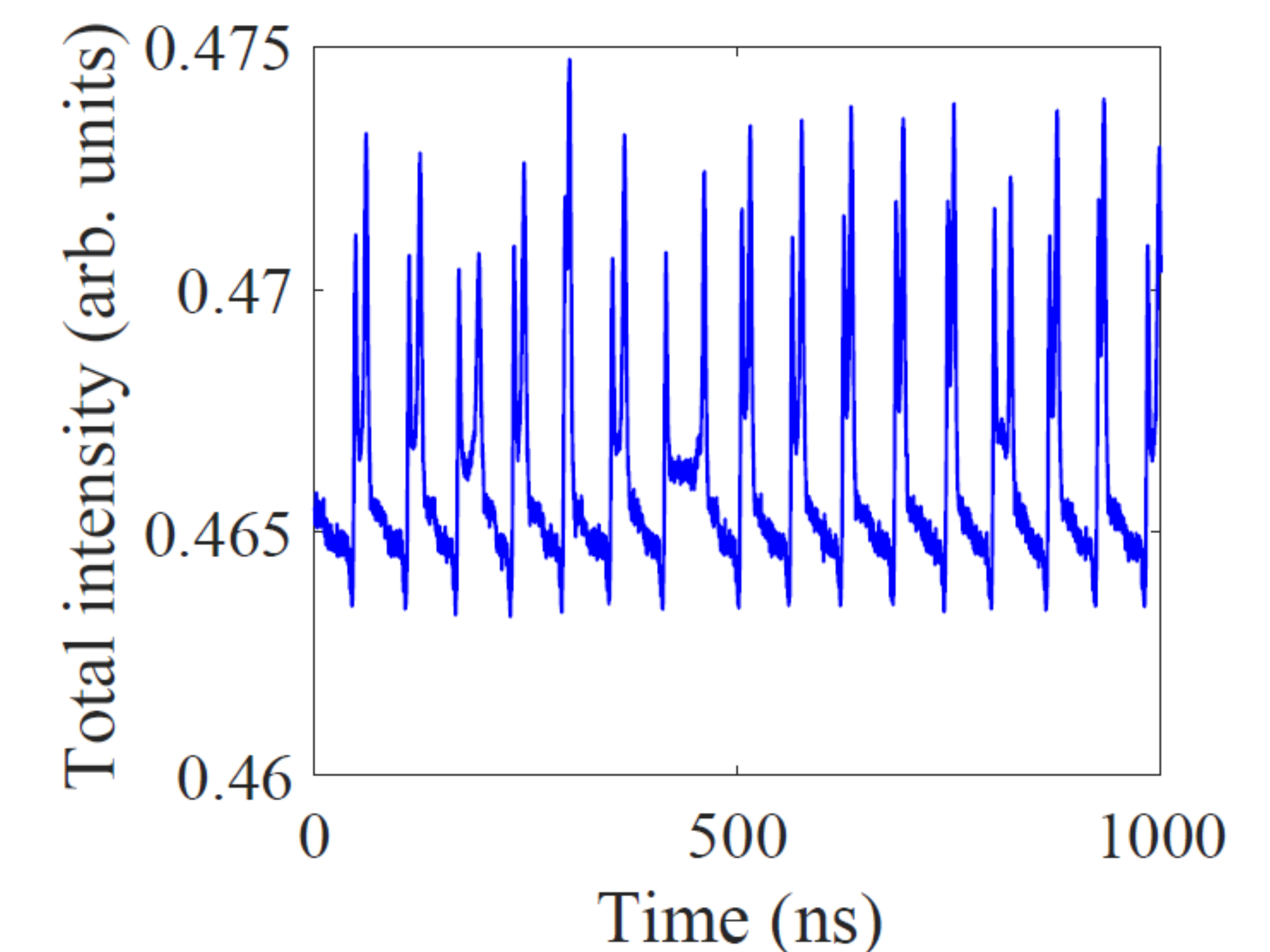}
\includegraphics[width=0.49\columnwidth]{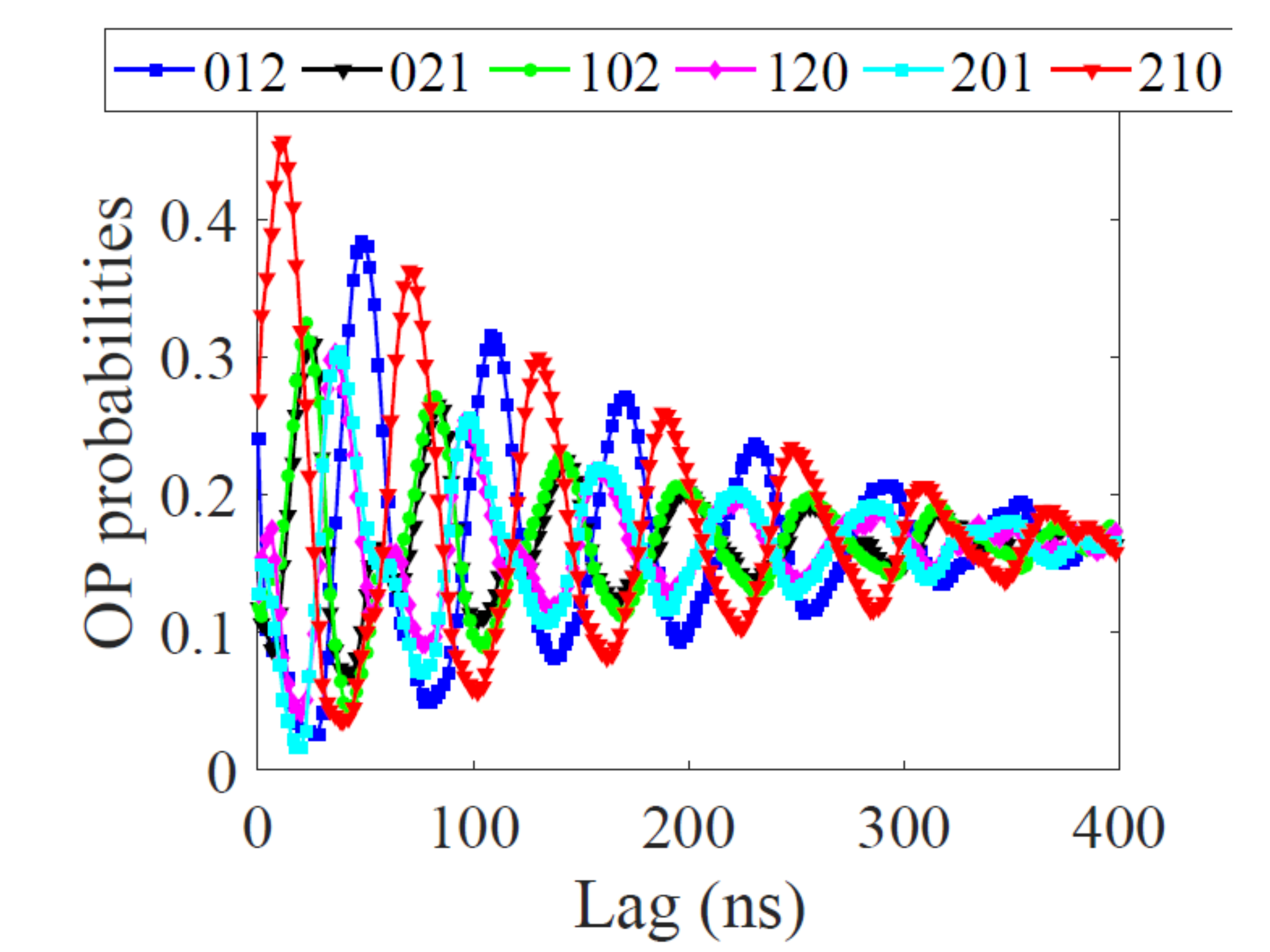}
\caption{Influence of the number of modes in the total intensity dynamics of the multimode model. Top: 8 modes, bottom: 16 modes.}\label{fig:mandel_16}
\end{figure}
\begin{figure}[htb]
\centering
\includegraphics[width=0.7\columnwidth]{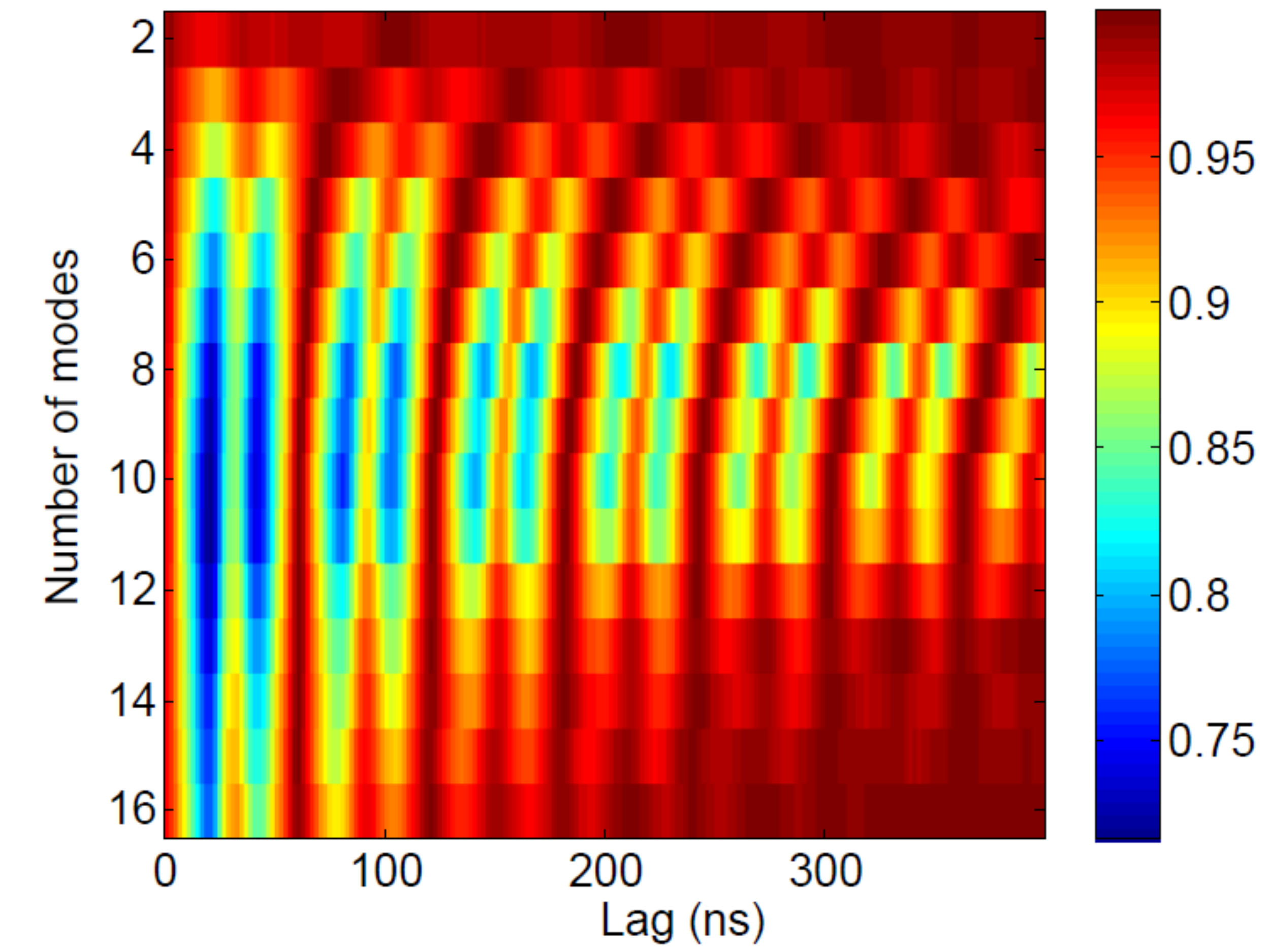}
\caption{Entropy computed from the ordinal probabilities in color code (normalized to the maximum entropy value, which corresponds to equally probable ordinal patterns) vs. the lag and the number of longitudinal modes.}\label{fig:entropy}
\end{figure}

\subsection{VCSEL model}

Here we simulate a well-known two-mode VCSEL model, which represents the competition of two linearly polarized modes. The model equations are \cite{martin}:

\begin{eqnarray}
\frac{dE_\pm}{dt}=k(1+i\alpha)(N\pm n-1)E_\pm - (\gamma_a + i\gamma_p)E_\mp, \nonumber \\
\frac{dN}{dt}= \gamma_N \left(J -N - (N+n)|E_+|^2- (N-n)|E_-|^2\right), \nonumber\\
\frac{dn}{dt}= -\gamma_s n -\gamma_N \left((N+n)|E_+|^2+(N-n)|E_-|^2\right).\nonumber
\end{eqnarray}

Here $E_\pm$ are the slowly varying amplitudes of the left and right circularly polarized components of the optical field ($E_\pm = E_x\pm iE_y$ with $E_x$ and $E_y$ being the orthogonal
linearly polarized components), $N$ is the total population difference between conduction and
valence bands, and $n$ is the population difference between the carrier densities with positive and negative spin values. $k$ is the field decay rate, $\gamma_N$ is the decay rate of the total carrier population, and $\gamma_s$ is the decay rate which accounts for the mixing of the populations with different spins. $\gamma_a$ and $\gamma_p$ represent gain anisotropies and birrefringence, respectively. $\alpha$ is the linewidth enhancement factor and $J$ is the normalized injection current.

\begin{figure*}
\centering
\includegraphics[width=0.67\columnwidth]{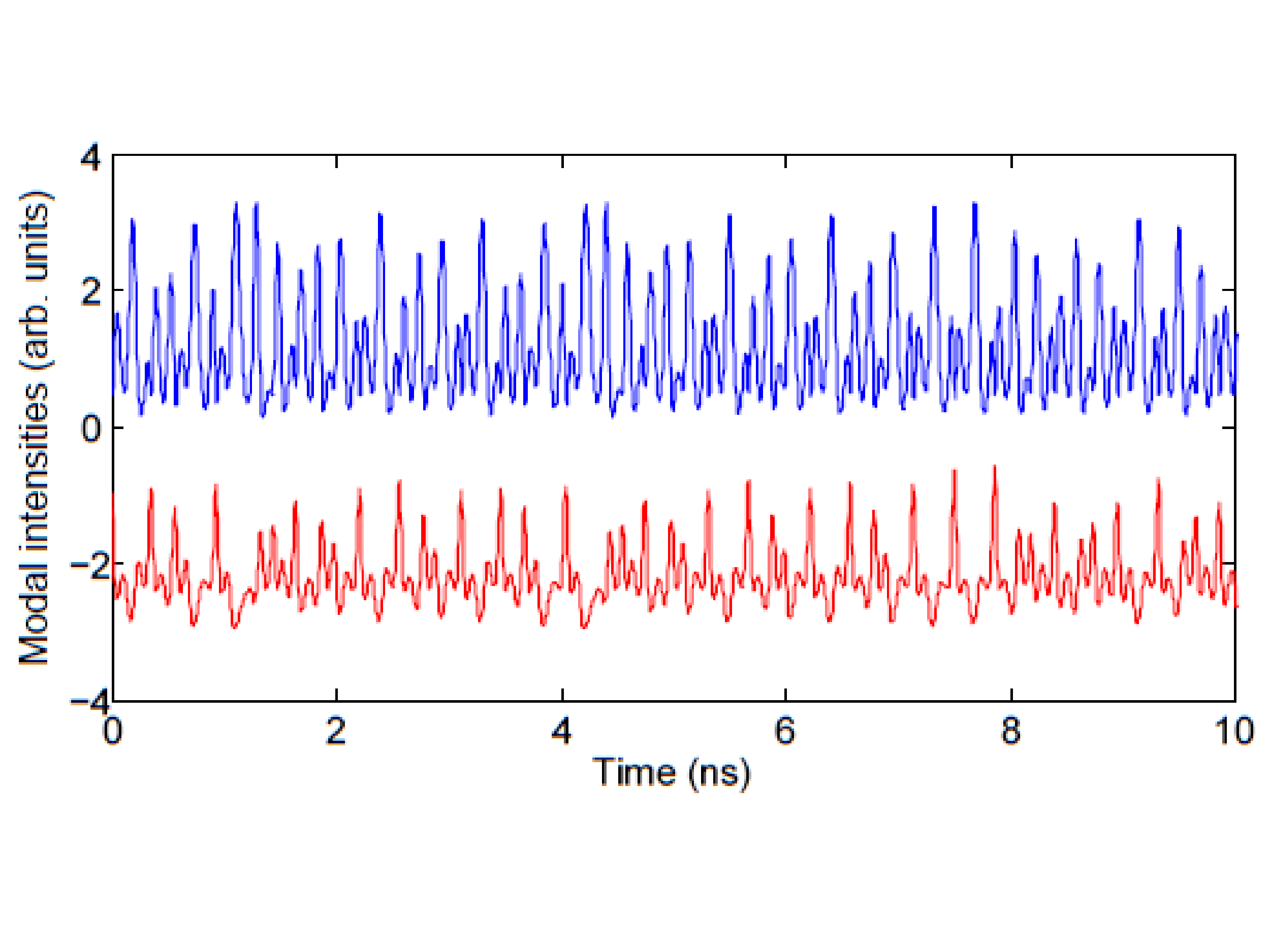}
\includegraphics[width=0.67\columnwidth]{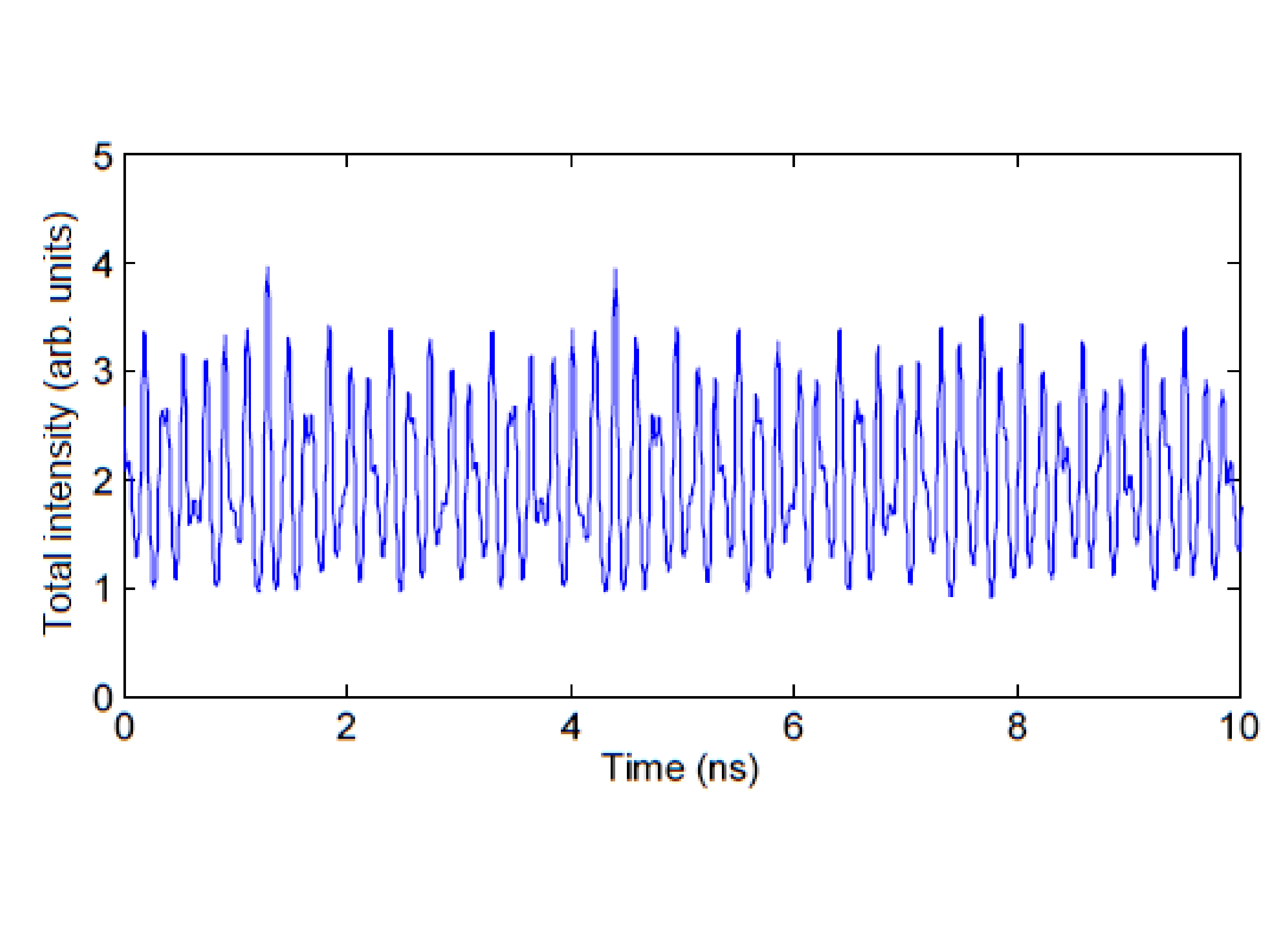}
\includegraphics[width=0.67\columnwidth]{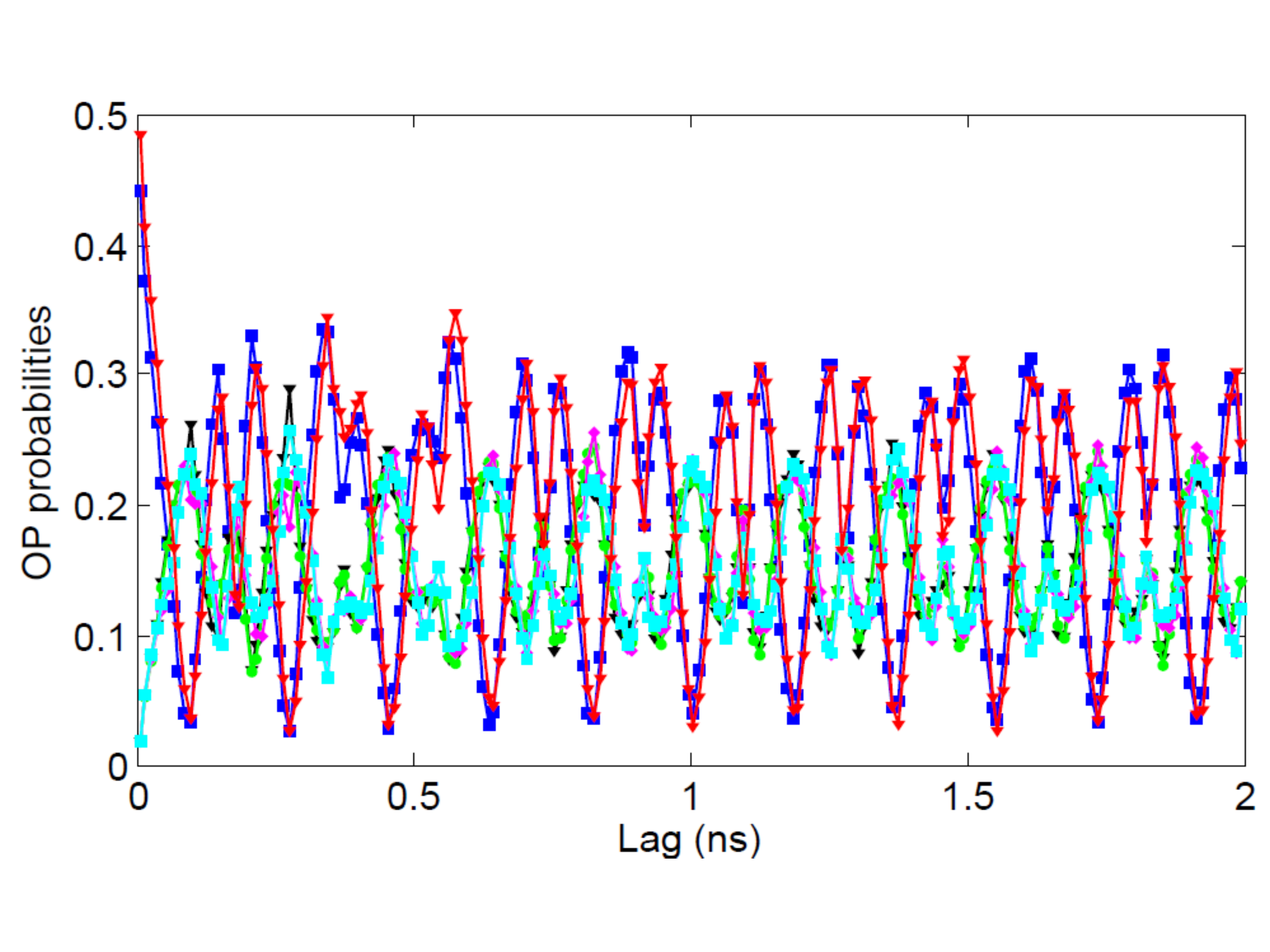}\\
\includegraphics[width=0.67\columnwidth]{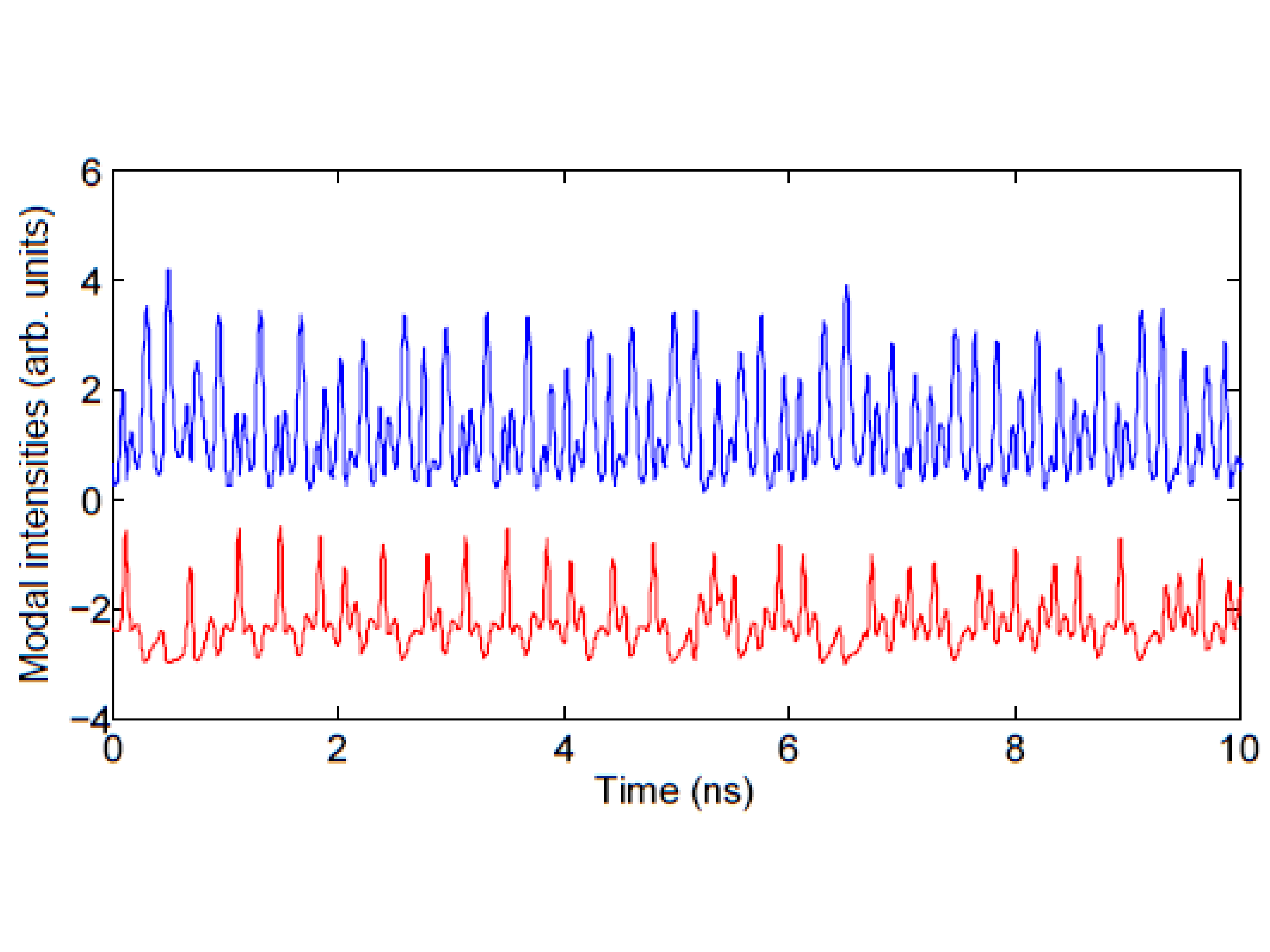}
\includegraphics[width=0.67\columnwidth]{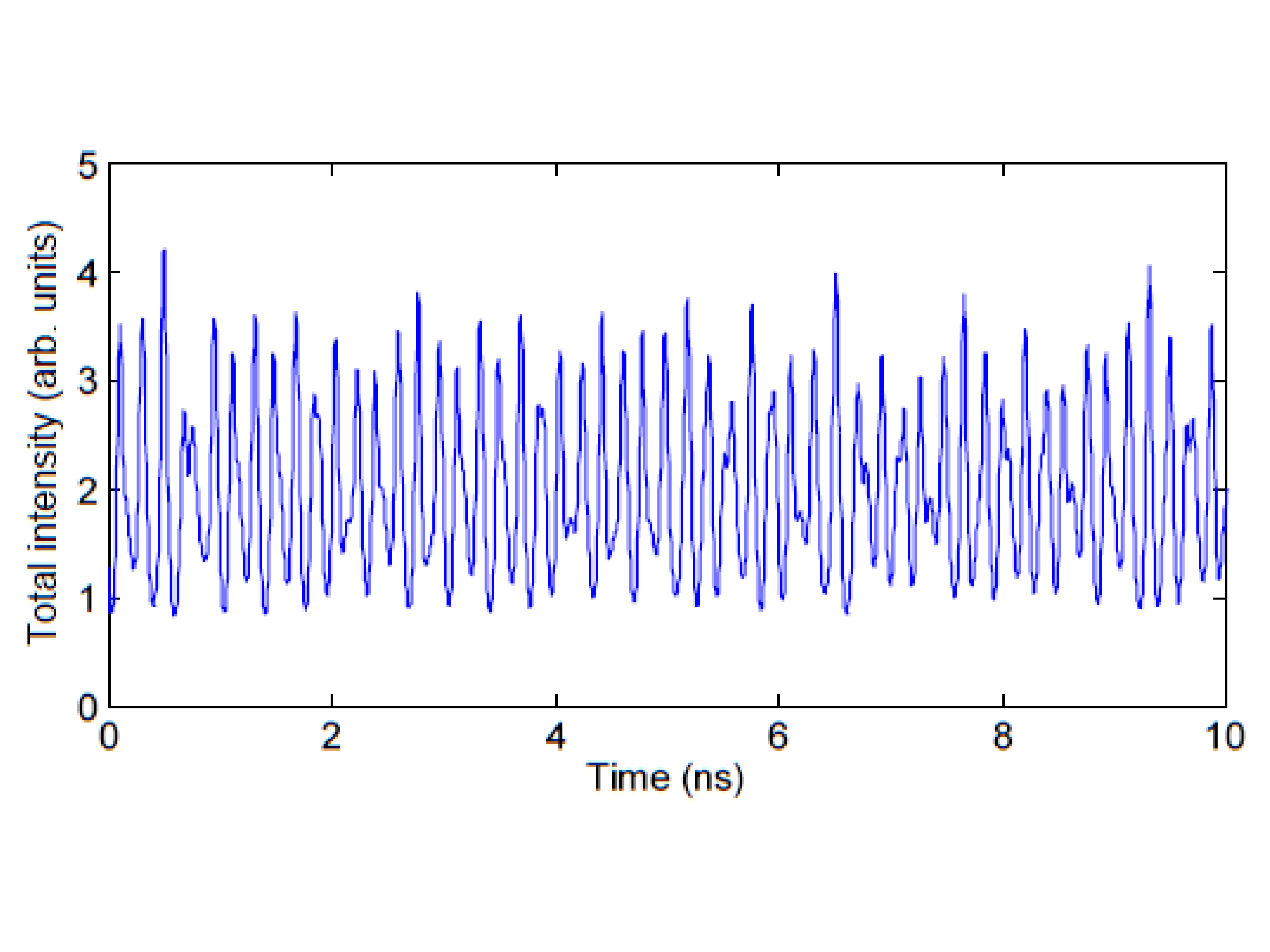}
\includegraphics[width=0.67\columnwidth]{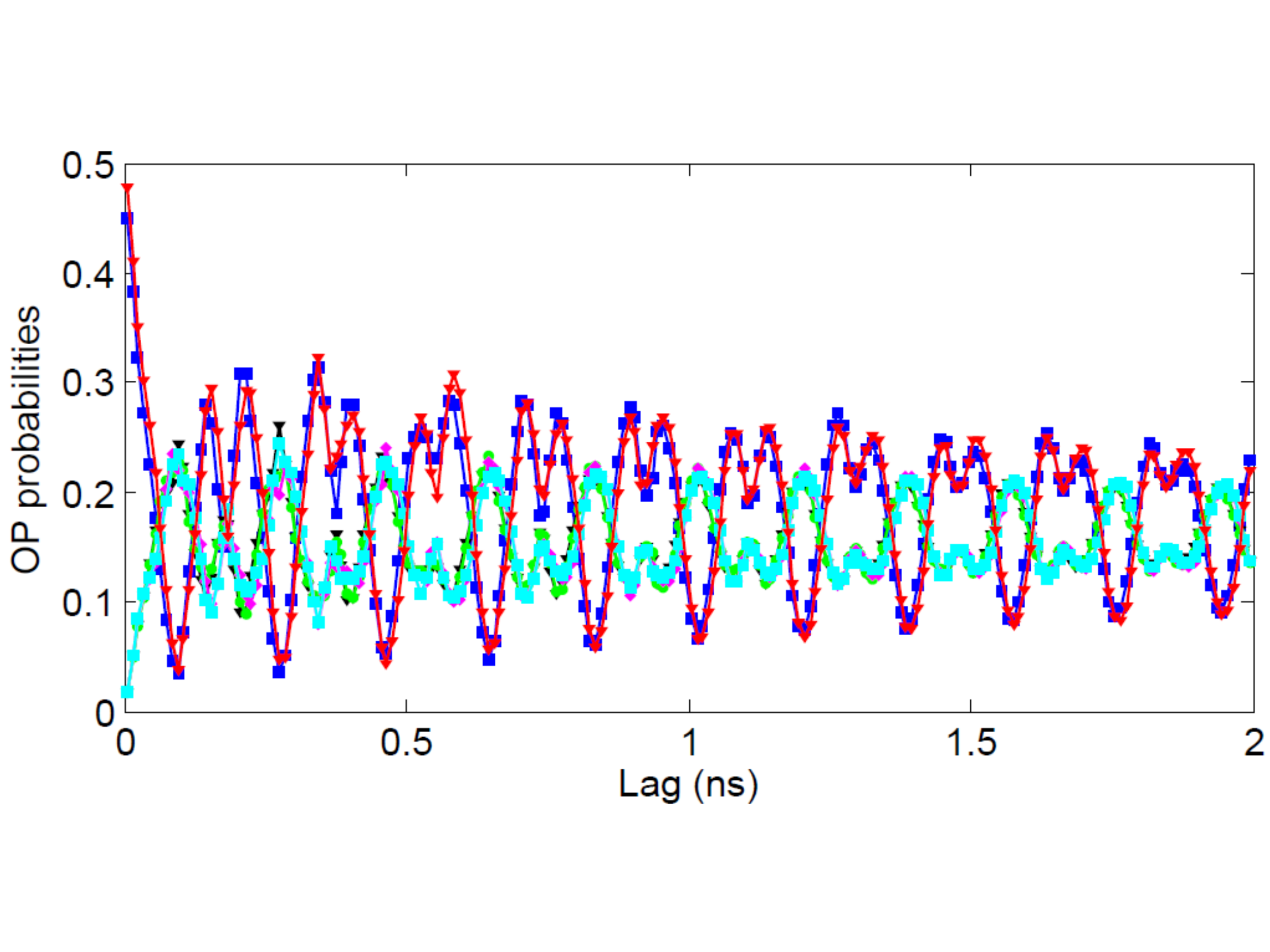}
\caption{Results of simulations of the VCSEL model with parameters $J=3$, $\gamma_a=-0.1$~ns$^{-1}$ and $\gamma_p=11.75$~rad ns$^{-1}$ (top row), $\gamma_p=12$~rad ns$^{-1}$ (lower row). The modal intensities (left panels) are shifted vertically for clarity. }
\label{fig:sfm}
\end{figure*}

We simulated the model equations with typical parameters, $k=300$~ns$^{-1}$, $\gamma_N=1$~ns$^{-1}$, $\gamma_s=50$~ns$^{-1}$, $\gamma_a=-0.1$~ns$^{-1}$ and $\alpha=3$, choosing $J$ and $\gamma_p$ such that both linear polarizations are unstable. Results are presented in Fig.~\ref{fig:sfm}, where we note that mode competition leads to an oscillatory dynamics in the total intensity, $|E_x|^2 + |E_y|^2$, which, when analyzed with ordinal patterns, has a similar variation of the ordinal probabilities with the lag, as that found in the main text. We present deterministic simulations but we have verified that the inclusion of spontaneous emission noise did not have any noticeable effect in the variation of the probabilities with the lag.

\newpage

\begin{thebibliography}{99}

\bibitem{Turitsyn2010} 
S. K. Turitsyn, S. A. Babin, A. E. El-Taher, P. Harper, D. V. Churkin, S. I. Kablukov, J. D.  Ania-Castanon,  V. Karalekas, and E. V. Podivilov, 
Nat. Photonics \textbf{4}, 231--235 (2010).

\bibitem{sergei_2017} S. V. Sergeyev, H. Kbashi, N. Tarasov, Yu. Loiko, and S. A. Kolpakov,
Phys. Rev. Lett. \textbf{118}, 033904 (2017).

\bibitem{real_time_2017}  K. Krupa, K. Nithyanandan, U. Andral, P. Tchofo-Dinda, and P. Grelu, 
Phys. Rev. Lett. \textbf{118}, 243901 (2017).

\bibitem{natphot} E. G. Turitsyna, S. V. Smirnov, S. Sugavanam, N. Tarasov,
X. Shu, S. A. Babin, E. V. Podivilov, D. V. Churkin, G. E.
Falkovich, and S. K. Turitsyn,
Nat. Photonics \textbf{7}, 783 (2013).

\bibitem{natcom_2015} D. V. Churkin, S. Sugavanam, N. Tarasov, S. Khorev, S. V.
Smirnov, S.M. Kobtsev, and S. K. Turitsyn,
Nat. Commun. \textbf{6}, 7004 (2015).

\bibitem{dima_2016}  A. K. Chattopadhyay, D. Nasiev, S. Sugavanam, N. Tarasov, and D. V. Churkin, 
Sci. Rep. \textbf{6}, 28492 (2016).

\bibitem{prl_2016} A. Aragoneses, L. Carpi, N. Tarasov, D. V. Churkin, M. C. Torrent, C. Masoller, and S. K. Turitsyn, 
Phys. Rev. Lett. {\bf 116}, 033902 (2016).

\bibitem{arecchi_1992} F. T. Arecchi, G. Giacomelli, A. Lapucci et al.,
Phys. Rev. A \textbf{45}, R4225 (1992).

\bibitem{gianni_1996} G. Giacomelli and A. Politi,
Phys. Rev. Lett. \textbf{76}, 2686 (1996). 
 
\bibitem{Lacasa2009} B. Luque, L. Lacasa, J. Luque, and F. J. Ballesteros, 
Phys. Rev. E \textbf{80}, 046103 (2009).

\bibitem{HVG} L. Lacasa and R. Toral, Phys. Rev. E \textbf{82}, 036120 (2010).

\bibitem{bandt_PRL_2002} C. Bandt and B. Pompe,
Phys. Rev. Lett. {\bf 88}, 174102 (2002).

\bibitem{review_rosso} M. Zanin, L. Zunino, O. A. Rosso, and D. Papo, Entropy 14, 1553 (2012).

\bibitem{review_amigo} J. M. Amigo, K. Keller, and J. Kurths, Eur. Phys. J. Spec. Top. 222, 241 (2013).

\bibitem{small_2014} X. Sun, M. Small, Y. Zhao and X. Xue,  Chaos 24 024402 (2014).


\bibitem{Mandelbrot_1968} B. B. Mandelbrot and J. W. Van Ness, SIAM Rev. \textbf{4}, 422 (1968).



\bibitem{Muzy1991} J. F. Muzy, E. Bacry, and  A. Arneodo, Phys. Rev. Lett. \textbf{67}, 3515 (1991).

\bibitem{Stolovitzky1994} G. Stolovitzky and K. R. Sreenivasan, Rev. Mod. Phys. \textbf{66}, 229 (1994).


\bibitem{perez_josa_2004} D. G. Perez, L. Zunino, and M. Garavaglia,
J. Opt. Soc. Am A \textbf{21}, 1962 (2004).

\bibitem{bio_2013} P. C. Bressloff and J. M. Newby, 
Rev. Mod. Phys. 85, 135 (2013).

\bibitem{olivares_2016} F. Olivares, L. Zunino, and O. A. Rosso,
Physica A \textbf{445}, 283 (2016).
 
\bibitem{npg_2017} J. M. Lilly, A. M. Sykulski, J. J. Early, and S. C. Olhede, 
Nonlin. Processes Geophys. \textbf{24}, 481 (2017).

\bibitem{ravetti} M. G. Ravetti, L. C. Carpi, B. Goncalves, A. C. Frery, and O. A. Rosso,
PlosONE \textbf{9}, e108004 (2014).

\bibitem{davidsen} T. Schweigler and J. Davidsen.
Phys. Rev. E \textbf{84}, 016202 (2011).

\bibitem{expressions} If $P=[p_i ; i=1, . . . ,M]$ is a discrete probability distribution,
with $M$ being the number of possible states, the Shannon entropy and the Fisher information are calculated as $S[P]=-\sum_{i=1}^{M} p_i\ln p_i$ and $F[P]=F_0 \sum_{i=1}^{M-1} \left[p_{i+1}^{1/2}-p_{i}^{1/2}\right]^2$ with $F_0$ a normalization constant \cite{ravetti}.

\bibitem{sfplane} C. Vignat and J. F. Bercher, 
Phys. Lett. A. \textbf{312}, 27 (2003).

\bibitem{BS_2007} C. Bandt and F. Shiha, J. Time Ser. Anal. \textbf{28}, 646 (2007).

\bibitem{libro_pik} A. Pikovsky, M. Rosenblum, and J. Kurths, \textit{Synchronization: A Universal Concept in Nonlinear Sciences} (Cambridge University Press, Cambridge, U.K., 2001).

\bibitem{prl_2016_2} D. Pierangeli et al., Phys. Rev. Lett. 117, 183902 (2016).

\bibitem{nat_comm_2017}  I. R. R. Gonzalez, B. C. Lima, P. I. R. Pincheira, A. A. Brum, A. M. S. Macedo, G. L. Vasconcelos, L. D. Menezes, E. P. Raposo, A. S. L. Gomes, and R. Kashyap, 
Nat. Commun. 8, 15731 (2017).

\bibitem{mandel} A. M. Yacomotti, L. Furfaro, X. Hachair, F. Pedaci, M. Giudici, J. Tredicce, J. Javaloyes, S. Balle, E. A. Viktorov, and P. Mandel, ``\textit{Dynamics of multimode semiconductor lasers}'', Phys. Rev. A \textbf{69}, 053816 (2004).

\bibitem{martin} J. Martin-Regalado, F. Prati, M. San Miguel, and N. B. Abraham, ``\textit{Polarization properties of vertical-cavity surface-emitting lasers}'', IEEE J. Quantum Electron. \textbf{33}, 765 (1997).

\end{thebibliography}
\end{document}